\documentclass[trackchanges]{aastex701} 
\usepackage{amsmath}

\begin{document}

\title[Toward realistic]{Toward a realistic test of black-hole movie correlations as a probe of extreme lensing}

\author[orcid=0000-0002-8599-4483]{Barbora Bezd\v{e}kov\'{a}}
\affiliation{Department of Physics, Faculty of Natural Sciences, University of Haifa, Haifa 3498838, Israel}
\affiliation{Haifa Research Center for Theoretical Physics and Astrophysics, University of Haifa, Haifa 3498838, Israel}
\email[show]{bbezdeko@campus.haifa.ac.il}

\author[orcid=0000-0002-6960-0704]{Shahar Hadar} 
\affiliation{Department of Physics, Faculty of Natural Sciences, University of Haifa, Haifa 3498838, Israel}
\affiliation{Haifa Research Center for Theoretical Physics and Astrophysics, University of Haifa, Haifa 3498838, Israel}
\email{shaharhadar@sci.haifa.ac.il}

\begin{abstract}
In an optically thin setting, gravitational lensing around a black hole creates multiple delayed images of any localized source, appearing at different positions on the observer's screen. 
Even if particular sources cannot be singled out, this effect may be revealed in a black-hole movie (time-dependent image)
by examining the two-point spatiotemporal correlation function of intensity fluctuations in the movie. Recently, the correlation function derived from a movie obtained by ray tracing of a general-relativistic magnetohydrodynamic simulation demonstrated the plausibility of this idea, even when the movie was blurred to angular resolutions corresponding to next-generation terrestrial very-long-baseline interferometric arrays. In this work, we further develop this method and study how some possible corruptions of the movie, motivated by real observational limitations, affect the ability to identify the extreme-lensing signatures by correlations. 
The limitations we investigate are movie resolution, cadence, temporal segmentation, duration, and noise.
The effects are compared for both the original and blurred movies. We discuss the implications of our analysis for upcoming observations of M87* and Sgr A*, the two main targets of the Event Horizon Telescope and its future extensions. Our case study suggests that the types of data corruption we have explored, in themselves, do not preclude an identification of extreme-lensing effects in correlation measurements.
We use our results to roughly estimate that, most importantly, for a successful correlation measurement, M87* may need to be continuously monitored for $\gtrsim 2$ years, while Sgr A* may need to be measured at a cadence $\gtrsim 1$ frame per minute.
Finally, we outline a new analysis method that reduces the data in the complete correlation function---a large dataset defined on a five-dimensional configuration space---to a two-dimensional histogram that could be useful for black-hole parameter inference.
\end{abstract}


\section{Introduction}
Despite the fact that black hole (BH) imaging has been theoretically discussed in the literature for decades, see, e.g., \cite{cunningham73,luminet79,viergutz93,falcke17}, the actual observational breakthrough came only several years ago when the first image of M87* was captured by the Event Horizon Telescope (EHT) collaboration using very long baseline interferometry (VLBI) \citep{EHT19I,eht19ii,eht19iii,EHT19IV,EHT19V,EHT19VI}. This capability opens new avenues for probing general relativity, as well as accretion physics, in the strong-field regime just outside the BH horizon.

Imaging a BH at the horizon scale allows one to directly observe the gravitational effect of a rotating (Kerr) BH on light propagation \citep[e.g.,][]{Carter1968,bardeen73}. In addition, there are environmental effects mostly arising from propagation in the surrounding plasma, which is often thought of as a geometrically thick and optically thin
disk accretion flow in the relevant frequency range \citep[e.g.,][]{rees82,reynolds96,EHT19V,lupsasca24}. Despite the remarkable achievement of the EHT collaboration, the current image resolution is yet insufficient to disentangle these two effects. 
A universal feature that, if measured, could provide a direct probe of strong-field general relativity
is the \emph{photon ring} --- a thin brightness enhancement of circle-like shape predicted to lie in BH images \citep[e.g.,][]{beckwith05,Johnson2020,perlick22,aratore24}. Key observable properties of a photon ring are determined mainly by the BH spin and inclination, and depend only mildly on the astrophysical parameters of the accretion flow 
\citep[e.g.,][]{ozel22,Vincent2022,jia24,Urso2025}. It is a consequence of light rays that undergo extreme gravitational lensing around the BH, completing several half-orbits $n$ before reaching the observer.
On the observer's screen, every source thus creates multiple images that converge exponentially (in $n$) toward a special closed critical curve
\citep{bardeen73}, defined as the boundary between the light rays that---when propagated back from the observer---reach infinity, and those which reach the BH horizon. 

The angular width of a photon ring is expected to be below $10~\mu$as \citep{BHEX2024} which is too small to be resolved by the current EHT (which has a spatial resolution of $\sim25~\mu$as) or the proposed next-generation EHT (ngEHT), with an expected resolution of $\sim15~\mu$as \citep{Doeleman2023,johnson23}. A possible way to overcome this limitation is to extend the VLBI observation into space, extending the longest baseline, and improving resolution. Proposed missions that are primarily motivated by the prospect of resolving the photon ring, such as the Terahertz Exploration and Zooming-in for Astrophysics \citep[THEZA;][]{Gurvits2022,Hudson2023} and the Black Hole Explorer \citep[BHEX;][]{BHEX2024}, are expected to reach a resolution $\sim 6~\mu$as.

An alternative approach to demonstrating the existence of a photon ring, which may not require such a fine-grained spatial resolution of the image, is the technique studied in the present paper. This technique is potentially viable even if the photon ring is not spatially resolved; it exploits the source's variability, combined with the BH's extreme gravitational lensing effects, to indirectly support the photon ring's existence.
Harnessing source variability to probe extreme-lensing effects was studied in, e.g., \cite{fukumura08,moriyama19,andrianov22,hadar23,harikesh25,zhang25}.
These methods exploit the fact that there is a time delay between images that come from the same source but correspond to different $n$. 

However, the effects of extreme lensing are not limited to time delays; images arising from rays that originate at the same source point, but correspond to different $n$, appear at different points on the observer's screen.
The sequence of multiple snapshots of BH images (``movie'') should thus contain spatiotemporal correlations: between different points on the screen, at particular time lags.   
Studying the two-point correlation function of intensity fluctuations as a function of both arriving time and positions on the observer's screen was suggested by \cite{hadar21}. They proposed that this quantity might be a suitable observable for detecting the extreme-lensing signatures without directly resolving the photon ring shape from time-averaged images. This suggestion was elaborated in \cite{bezdekova26}, where the two-point correlation function of intensity fluctuations was calculated from a BH movie derived from a general relativistic magnetohydrodynamics (GRMHD) numerical simulation. 
The study of correlations revealed extreme-lensing signatures, even in movies with resolutions of (next-generation) terrestrial-VLBI missions such as ngEHT. Except for the study of correlations at different blurrings (resolutions), many other realistic limitations of the movie were not discussed in this work. In addition, two-point correlation functions in the visibility domain were studied in \cite{wong24}, showing lensing signatures at long baselines.

Can extreme-lensing signatures be measured using correlations in upcoming VLBI observations? In which sources and instruments is there a better prospect of delivering such a detection, and are there particular inputs that we can provide now, at the planning stage of upcoming missions, that could boost the chances of making the detection?
The main aim of this work is to expand and advance our investigation of these questions. 
Addressing them required us to artificially corrupt the data in various ways, simulating different realistic limitations: cadence and duration constraints, temporal segmentation, and noise injection. We also developed new analysis techniques to interpret the results for the correlation function.
While in \cite{bezdekova26} we demonstrated that the suggested technique is an effective analysis tool for a simulated movie blurred to realistic resolution, here we provide extended evidence in support of the hypothesis that extreme-lensing signatures may be identified, using correlations, in real measurements. 
However, more work is required to firmly establish the effectiveness of movie correlations for realistic BH movie analysis. One significant inherent limitation that we leave for future investigation is the effect of finite $(u,v)$-coverage, or in other words, the fact that BH images are created using signals from a limited number of separated antennas. In addition, we must broaden the region of parameter space (BH spin and inclination, and accretion parameters) that we explore with our computations. The final section of the paper discusses these prospects.

\section{Simulated black hole movie}  \label{sec:simulated movie}
The movie analyzed in this work is the same as the one used in \cite{bezdekova26}. Here we briefly summarize its main properties. It originates from a 3D GRMHD simulation of a magnetically arrested disk (MAD) produced by {\tt{iharm3d}} (resolution of $384\times192\times192$) as described in \cite{prather21} which was ray traced with {\tt{ipole}} \citep{moscibrodzka18}. An ideal fluid was assumed and a thermal electron distribution was prescribed, with ion-to-electron temperature ratio parameterized by $R_{\rm low} = 1$ and $R_{\rm high} = 40$ \citep{Moscibrodzka2016}. Images were obtained assuming an M87*-like source, i.e., a BH with mass $M=6.5\times10^9 M_\odot$, spin $a_*=0.9375$, and inclination angle between the spin axis and the line of sight $\theta_o=163^\circ$. The source was assumed to be located at a distance $D = 16.9$ Mpc. The total duration of the movie is $4000 \, GM/c^3$, returning images at a cadence of $0.5 \, GM/c^3$. 

Most of the results presented in this work were obtained from a movie generated with a \textit{slow-light} ray-tracing prescription. This corresponds to a realistic situation where both light propagation and the fluid state evolve simultaneously within the simulation time, along the entire extent of the light ray in the simulation domain.
In Sec.~\ref{sec_n0} we applied an additional ray-tracing prescription, where only rays propagating directly towards the observer (along the segments labeled as $n=0$, see below) were considered. The movie thus does not include extremely-lensed photons, that undergo one or more half-orbits $n$ around the BH. The segment boundary was defined at turning points of $dz/ds$, where $z=r \cos \theta$ with $r$, $\theta$, being the Boyer-Lindquist coordinates, and $s$ denotes the affine parameter. 
These turning points correspond to the reversal of geodesic vertical propagation, where it starts traveling back toward the $z=0$ midplane. 
Thus, correlations stemming from gravitational lensing effects cannot be detected in the direct-only ($n=0$) movie, as these correlations are a consequence of multi-path light propagation, as will be explained in Sec.~\ref{sec_comments}.
Even though the direct-only prescription is not completely realistic, it allows us to identify the main features that arise from extreme-lensing correlations. It may also be thought of as roughly mimicking the effect of strong absorption, which suppresses higher-order images.

Another crucial aspect of the movie is its spatial resolution. The original movie we analyzed has a resolution (pixel size) of $0.5~\mu$as, where for a BH at distance $D$ given above, one has $\approx 3.8~\mu$as per $GM/c^2$. Our original resolution is thus not achievable with any of the upcoming VLBI instruments.  
To mimic the realistic resolutions of upcoming observations, we blurred the images by convolution with Gaussian kernels of various characteristic length scales, defined by their full width at half maximum (FWHM). This allowed us to study the effect of blurring on the different aspects of the correlation function described in the paper. We devote Sec.~\ref{sec_blur} to investigate blurring effects.

\section{Some comments about the two-point correlation function} \label{sec_comments}
Following the notation introduced in \cite{bezdekova26}, the normalized (continuous) two-point correlation function is given as
\begin{equation}\label{corr_defT}
\mathcal{C}(T,x,y,x',y')=\frac{1}{\sigma\left[\Delta I(t,x,y)\right] \sigma\left[\Delta I(t,x',y')\right]}\langle \Delta I(t,x,y)\Delta  I(t-T,x',y')\rangle,
\end{equation}
where $\Delta I(t,x,y)=I(t,x,y)-\langle I(t,x,y) \rangle$ denote intensity fluctuations defined in the (asymptotic) observer time and Cartesian, Bardeen screen coordinates \citep{bardeen73} --- $(t,x,y)$. The temporal standard deviation of the intensity at the screen point $(x,y)$ is denoted by $\sigma\left[\Delta I(t,x,y)\right]$, and $\langle \rangle$ represents an average over all available times. Generally, $\mathcal{C}$ is not symmetric under time reversal. In our study, we use the convention (without loss of generality) that $T\geq0$ and that the intensity at the point $(x',y')$ is measured earlier than the intensity at the point $(x,y)$, as indicated in Eq.~\eqref{corr_defT}.  
Note that, by definition, $\mathcal{C}(0,x,y,x,y)=1$.

The definition~(\ref{corr_defT}) assumes that the signal is continuous; real observations deal with discrete signals, 
where the correlation function can be defined as \citep[cf.][]{edelson88}
\begin{equation} \label{corr_def_discr}
  \mathcal{C}(T,x,y,x',y')=\frac{1}{N_T}\sum_{i}\frac{\Delta I(t_i,x,y)\,\Delta I(t_i-T,x',y')}{\sqrt{\left(\sigma^2[\Delta I(t_i,x,y)]-e^2_{(x,y)}\right)\left(\sigma^2[\Delta I(t_i,x',y')]-e^2_{(x',y')}\right)}},
\end{equation}
where $N_T$ denotes the number of data pairs contributing to the lag $T$; $I(t,x,y)$ is a discrete array,
and $e_{(x,y)}$ denotes the estimated measurement errors. For our data set, we have $e_{(x,y)}=0$, but this formula can easily be adapted for more complicated situations.

Black-hole movies are shaped both by the complex astrophysical processes that emit the radiation and by the comparatively simpler gravitational lensing that governs how this radiation propagates to the observer.
The emission physics depends on many parameters and is generally computationally costly to model, while gravitational lensing effects are universal---depending only on BH mass and spin, and on observer inclination---and analytically tractable.
In principle, the two-point correlation function provides a means of distinguishing between properties of the emitting plasma accreted into the BH, and properties of the spacetime geometry, even when the photon ring is not spatially resolved in the time-averaged image \citep{hadar21}. 

We expect the two-point correlations of physical properties of the accretion flow, which are functions of the spacetime position (e.g., velocity, temperature, density,~$\ldots$) to be local---i.e., to peak in the case where the two spacetime points coincide and monotonically decrease as they are separated. 
In contrast, extreme-lensing effects induce non-local correlation on the observer screen, between pairs of points separated both spatially and temporally. Their existence is manifested as a local maximum of the correlation function at $\left( T>0, (x,y)\ne (x',y') \right)$, which was demonstrated by \cite{bezdekova26}.

Non-local correlations in BH movies arise as a result of extreme gravitational lensing; namely, due to the effect of multi-path propagation. A source fluctuation occurring at any spacetime event is connected to the observer via multiple distinct null geodesics, that may be labeled by the number of half-orbits $n$ they undergo around the BH.\footnote{At high enough spin, inclination, and half-orbit number $n$, there exist distinct null geodesics that wind around the azimuthal direction a different number of times; in those cases $n$ must be supplemented with an azimuthal winding number in order to uniquely specify the null geodesic \citep{zhou25}.} 
Photons traveling on light rays with different $n$ arrive at the observer's screen at different positions and different times, with higher-$n$ photons lagging behind.  
In the small-inclination case, relevant for our situation, the expected time lag between different sequential images of the same point source changes only mildly with BH spin. Moreover, for approximately equatorial sources at small inclination, the expected time lag is $\sim 15\,GM/c^3$ \citep[e.g.,][]{gralla20lensing}. 

Given that the flux density of the $n^{th}$ image (made up of photons that underwent $n$ half orbits) decreases exponentially with $n$ \citep{Gralla2019,Johnson2020}, the most relevant contribution to the correlation function is expected to be between the $n=0$ and $n=1$ images. Hence, in the present work, we focus on the $n=0,1$ images when interpreting our results. We note that, depending on the application, estimates about the mutual behavior of photons with different $n$ often use the so-called near-photon shell approximation $n\gg1$, which provides enhanced analytical control. However, for rays with $n=0,1$ this approximation is not sufficiently precise for our needs (even though for many cases it provides a surprisingly decent estimate, since it is a rapidly converging approximation). We therefore resort to numerical evaluation of $n=0,1$ pairs of images of equatorial point sources; see details in Sec.~\ref{sec_stat}.

In general, $\mathcal{C}$ \eqref{corr_defT} 
is defined over a 5D domain: the configuration space $(T,x,y,x',y')$. Its high dimension makes it challenging to visualize the full correlation structure and study the dependence on all variables simultaneously.
One way to reduce the dimension is to integrate over some directions of configuration space, obtaining a \emph{coarse-grained} correlation function. For example, integrating over all spatial screen positions $(x,y,x',y')$ gives the light-curve autocorrelation function, which depends only on the time lag $T$. It was recently studied in real \citep{wielgus22} and simulated \citep{cardenas24,bezdekova26} data, and was found to be too coarse-grained to show signatures of extreme-lensing effects. Another example was studied in \cite{hadar21}, where the general 5D correlation function was first defined. An observable that was computed there analytically, in the $n \gg 1$ approximation, was a partially coarse-grained version of $\mathcal{C}$ in which one integrates over both radial positions of the two points, but does not integrate over both angles on the screen.
However, as we discussed in \cite{bezdekova26}, any type of coarse graining will generically lose information about the correlation structure. In case the data is insufficient to construct the full 5D function, it may well be useful; but if sufficient data does exist, coarse graining does not introduce any advantage so we will avoid it whenever possible.

Nevertheless, for visualization purposes, we somehow need to reduce the dimension of the domain over which we present the correlation function. We therefore often fix the spatial position of the point $(x',y')$ to some constant $(x_0',y_0')$. When doing this, we denote by $\mathcal{C}=\mathcal{C}(T,x,y)$ the correlation function restricted to the relevant 3D domain. We can then evaluate many such functions for many fixed-point pairs $(x_0',y_0')$, ideally covering the full observer screen. In theory, therefore, this approach does not lose any information compared to the original function $\mathcal{C}(T,x,y,x',y')$.
We note that while below our results (up to Sec.~\ref{sec_stat}) are shown for a single fixed point, $(x_0',y_0')=(-1.317,3.819)~GM/c^2$, our findings were studied and verified for a larger set of points to ensure that they are general enough. As demonstrated in the supplementary material of \cite{bezdekova26}, the choice of fixed point can be rather arbitrary---anywhere on the observer screen---and the lensing peak is still identifiable. Nevertheless, whenever relevant in the remainder of the paper, we comment on the dependence of the discussed effects on the position of the fixed point.

\section{A comparison of complete (all \lowercase{$n$}) and direct-only (\lowercase{$n=0$}) correlation maps}\label{sec_n0}
\begin{figure}[h!t]
\centering
\includegraphics[width=.95\textwidth]{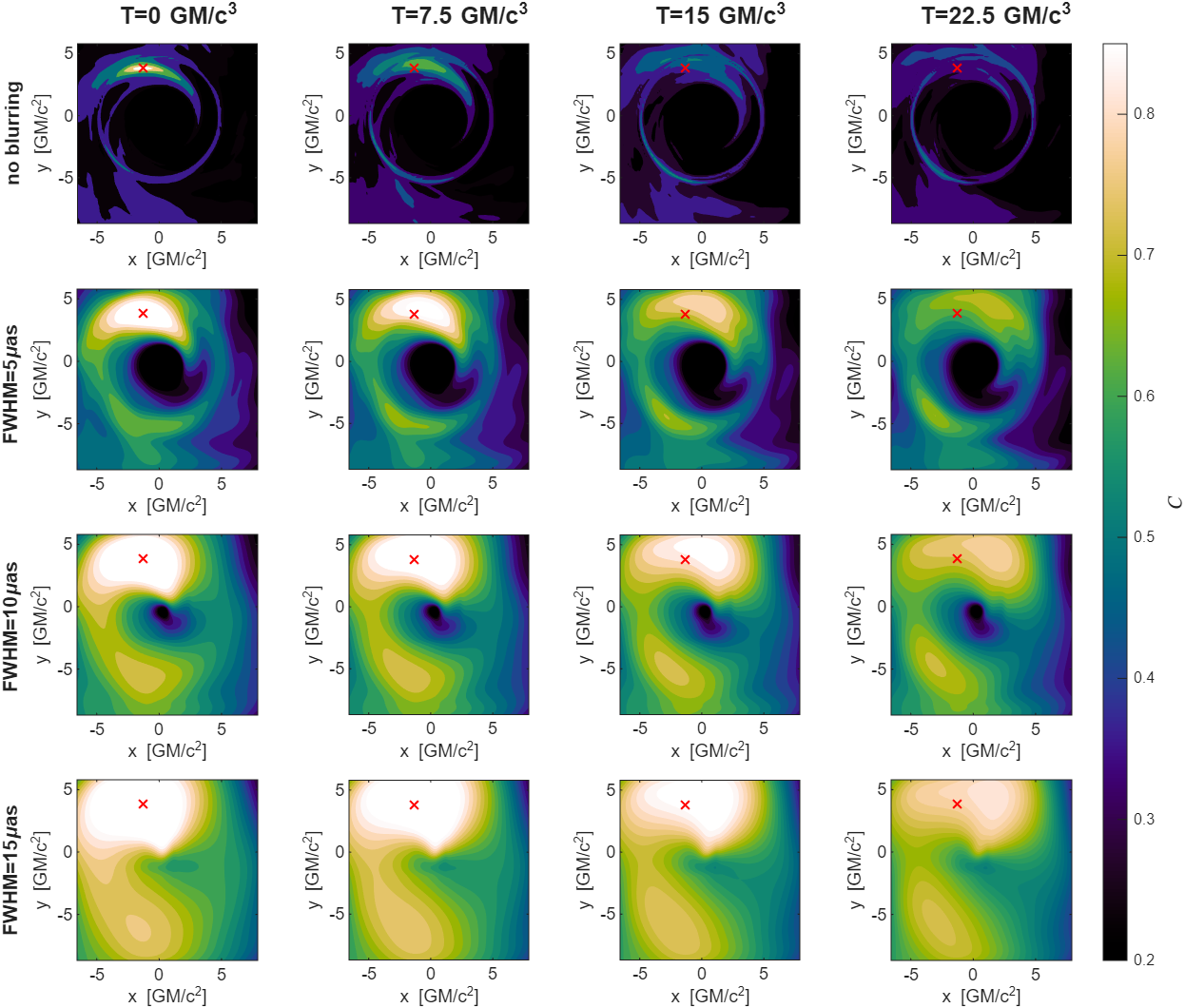}
\caption{Two-point correlation maps $\mathcal{C}(T,x,y,x_0',y_0')$ (see Eq.~\eqref{corr_defT} for the definition), presented as a function of the screen coordinates of the later point $(x,y)$ at different values of the time lag $T$ (columns) for movies with various blurring kernels (rows) obtained from the full slow-light movie. The fixed (earlier) point, marked by red crosses, is located at  $(x_0',y_0')=(-1.317,3.819)~GM/c^2$.}\label{T_n1}
\end{figure}

In this section we will present results for the two-point correlation maps $\mathcal{C}(T,x,y,x_0',y_0')$, for the fixed point choice $(x_0',y_0')=(-1.317,3.819)~GM/c^2$. We contrast two different cases: correlation maps of complete movies, where the contributions of light rays of all $n$ are included, and direct-only correlation maps, which are computed from the $n=0$ part of the movies only. 
The correlation maps were computed from the unblurred movie and three blurred ones, namely with FWHM=5, 10, and 15$~\mu$as.
Two-dimensional fixed-$T$ slices of these maps are shown in Fig.~\ref{T_n1},
for the time lags $T$=0, 7.5, 15, and 22.5~$GM/c^3$. Generally speaking, the maps show an approximate circle of enhanced correlation, roughly coinciding with the critical curve at a radius $\sim 5~GM/c^2$. On this circle, there is an apparent correlation enhancement opposite the position of the fixed point at time lags around $T$=15~$GM/c^3$. This can be seen for all blurrings, however in the blurred movie with FWHM=15$~\mu$as the maximum of the lensing peak is reached at shorter lag, see Sec.~\ref{sec_ident}. In the unblurred case (the first row of Fig.~\ref{T_n1}) the maximum lensing correlation is also present at a time lag around $T$=15~$GM/c^3$, but it is
spatially narrow and its correlation is not as high as in the blurred movies. In addition to the lensing lobes, located in the bottom middle/left part of the maps, the significant structure of a thin ring is apparent, especially in the blurred cases. The ring becomes wider for larger blurrings. 
The appearance of the ring confirms that the ray-traced movie includes rays with $n>0$.

\begin{figure}[h!t]
\centering
\includegraphics[width=.95\textwidth]{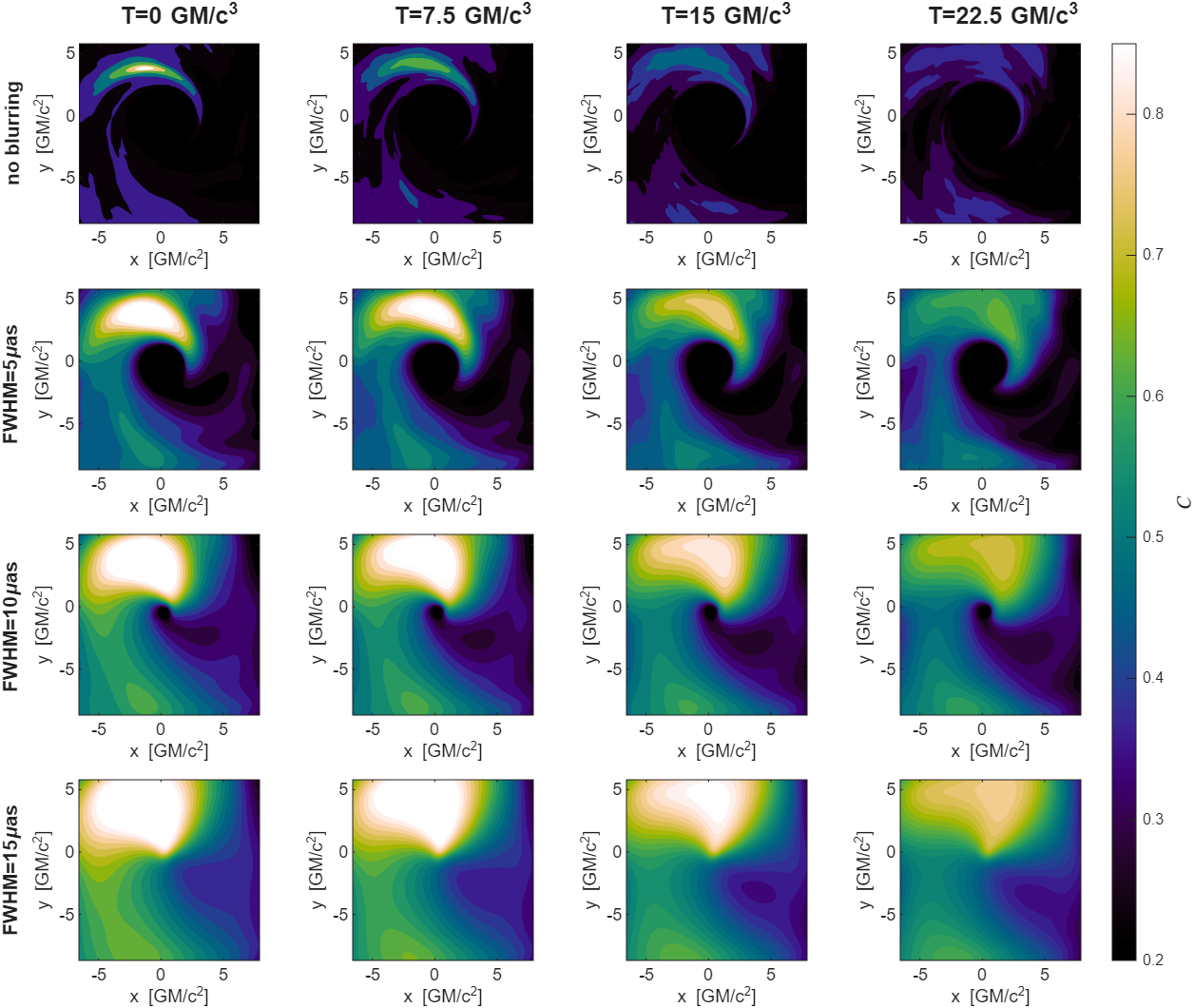}
\caption{Direct-only correlation maps, computed using the $n=0$ movie which omits the extreme-lensing contributions. The conventions, as well as the value of the fixed point, are the same as in Fig.~\ref{T_n1}.}
\label{T_n0}
\end{figure}

In contrast,
the direct-only correlation maps, computed from the $n=0$ movie and shown in Fig.~\ref{T_n0}, show no apparent ring structure.
Moreover, the correlation at the positions where the lensing peaks appear in the complete maps is significantly lower compared to the same regions in Fig.~\ref{T_n1}.
Furthermore, in these regions the correlation is maximal around $T$=0~$GM/c^3$ and decreases as the lag $T$ increases, without manifesting any substantial increase at later $T$, in contrast to its behavior in Fig.~\ref{T_n1}.

The comparison performed in this section aims to further elucidate the signatures of extreme lensing in correlation maps. 
Understanding how to identify these universal signatures and use them to extract information on the observed BH's geometry is our main goal. 
However, in order to investigate whether these features could be distinguished in upcoming VLBI campaigns, more simulated sources of data corruption must be introduced. In the following sections we further develop our techniques to identify extreme lensing features, simulate various forms of data degradation beyond image blurring, and investigate the detectability of these features in the (controllably) degraded data.

\section{Identification of the extreme-lensing correlation peak}\label{sec_ident}
The next step following the calculation of the full correlation function 
\eqref{corr_defT} is to analyze and interpret the result. In this section we describe how we
identify correlation peaks across the full configuration space, in order to facilitate this analysis. As was demonstrated in \cite{bezdekova26}, 
a useful way to reduce the dimensionality of the configuration space is to fix the (early) point to some constant value, $(x',y')=(x_0',y_0')$, and study the correlation structure in the remaining three dimensions. Subsequently, this procedure will be repeated for many different values of the fixed point $(x_0',y_0')$.
Here we focus on how to find the lensing correlation peak in the reduced 3D space spanned by $(T,x,y)$. 

From the general definition \eqref{corr_defT}, for any $(x_0',y_0')$, we expect the extreme-lensing peak to appear at $T>0~GM/c^3$ at some $(x,y)\ne(x_0',y_0')$. If, for simplicity, one assumes an equatorial source, given the BH spin and inclination angle, the precise position of the extreme-lensing peak on the observer screen can be theoretically calculated. This provides a good estimate for where to preferably search for the lensing peak, and as a starting point for our analysis.

Alongside the extreme-lensing peak, we know that there will be a primary correlation peak, arising mainly from local correlations in the astrophysical flow, which occurs at $T=0~GM/c^3$ and $(x,y)=(x_0',y_0')$. In fact, by definition, the correlation function assumes the maximum possible value at the coincidence point, $\mathcal{C}(0,x_0',y_0',x_0',y_0')=1$. This correlation peak will decrease as $T$ increases, while the lensing peak will increase. 
One may naively expect that, at some fixed $T>0$,
the lensing correlation peak will be a global maximum in the $(x,y)$-plane, and in particular that  
at that $T$ there exists some $(x,y)$ for which the correlation is higher than at $(x_0',y_0')$. 
It turns out that within the range of analyzed $T$ this is not always the case, especially for blurred movies; this may be seen in several correlation maps in this paper. For this reason, in order to find the position of the lensing correlation peak, 
one must search for \emph{local} correlation maxima; we first search for local maxima in the $(x,y)$-plane for various values of $T$, and then locate the peak in the full space spanned by $(T,x,y)$, where $T>0~GM/c^3$.

\begin{figure}[h!t]
\centering
\includegraphics[width=.65\textwidth]{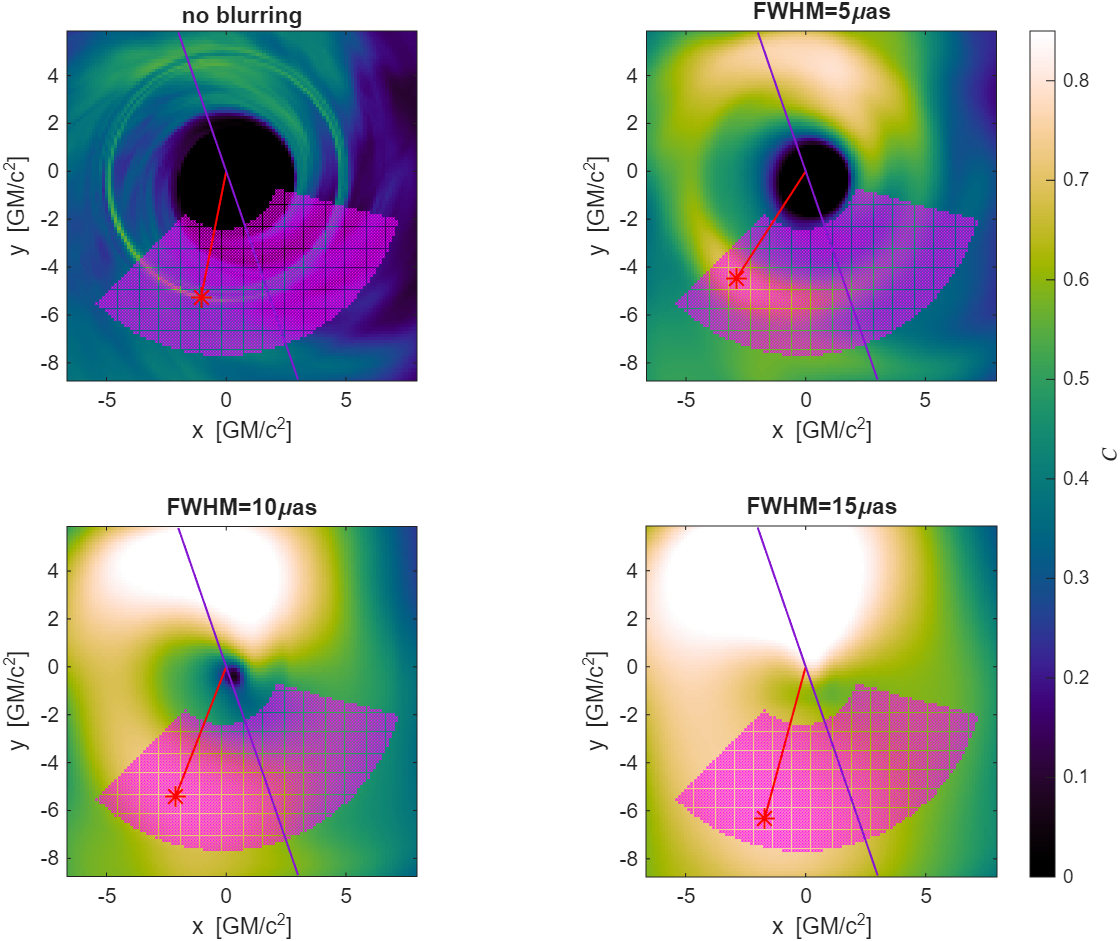}
\caption{Range of observer screen coordinates (magenta) over which we search for maximal correlation (red star), corresponding to the lensing peak. The purple line connects the fixed point $(x_0',y_0')$ and the origin, where it intersects the red line. 
$\delta\varphi$ is defined as the angle between the purple and red lines.}\label{ident_ex}
\end{figure}

Due to the dominance of the astrophysical peak, we reduce the range of $(T,x,y)$ over which we search for the local lensing peak. 
Assuming general relativity and optically thin emission, for modest inclinations, the lensing peak should appear around $T\sim 15~GM/c^3$ \citep[e.g.,][]{gralla20lensing}, with a weak dependence on BH spin. On the lag axis, we searched for the peak in the range $0<T<40~GM/c^3$. 
Regarding the screen coordinates $(x,y)$, 
if we additionally assume small $a_*$ and mild inclination, we expect the lensing peak to appear rotated by $\sim 180^\circ$ with respect to the center of the image,\footnote{which can be defined as the weighted average of the pixels composing the empirical critical curve, see \cite{bezdekova26}.} and quite close to the critical curve.
The dominant lensing effect of general BH spin $a_*\neq0$, even at mild inclination, is expected to be an additional frame-dragging effect that will vary the rotation angle of the lensing peak relative to the astrophysical peak, allowing it to differ from $\sim 180^\circ$.

Since a priori (in measurements) we would not know how far the departure from this expectation is, we have to make a choice for the search region which would best accommodate several constraints.
The most significant limitation is the extension of the astrophysical peak. Since its effect expands for larger blurrings across the $(x,y)$-plane, the search region cannot be too extended in the angular domain. For general inclination, this 
might
result in missing some of the lensing peaks since their position could be too close to the astrophysical peak.
If so, the routine---which picks out the point of maximum correlation in the specified region---will
pick out points that lie on the edge of the region. For this reason, we add the requirement that the position of the 
lensing peak lies in the bulk of the specified region. Peaks detected on the edges are discarded in the present analysis.

Some examples of search regions for lensing peaks, on top of correlation maps of different blurrings, are shown in magenta in Fig.~\ref{ident_ex}. The fixed point in these maps is the same as specified in Sec.~\ref{sec_n0}, and the lag is fixed such that the correlation value at the lensing peak is maximal. The position of maximal correlation detected in the specified region, over all lags in the search, $T=[0,40]~GM/c^3$, is plotted by red stars. The angle between the purple and red lines indicates the deviation of the peak from a $180^\circ$ rotation relative to the astrophysical peak; we denote this angle by $\delta \varphi$, where $\delta \varphi>0$ if the deviation is clockwise.
Comparing $\delta \varphi$ in the different maps provides a measure of stability of our results as a function of blurring.

\begin{figure}[h!t]
\centering
\includegraphics[width=.65\textwidth]{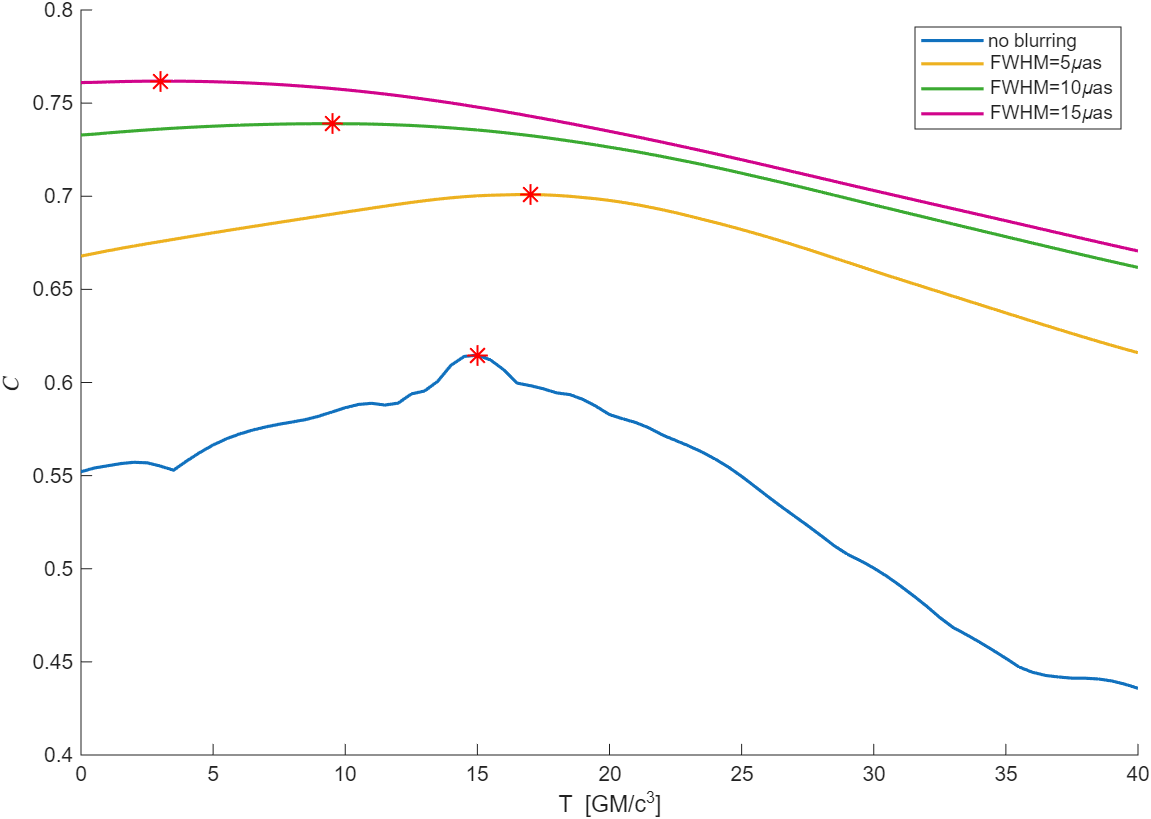}
\caption{Maximal value of the correlation in the search regions defined in Fig.~\ref{ident_ex}, as a function of the lag $T$, for movies of different blurrings. The temporal position of the absolute maximum in a given area is marked by a red star.}\label{max_corr_ex}
\end{figure}

As another check, and in order to locate the lensing peak across the remaining configuration-space coordinate $T$, Fig.~\ref{max_corr_ex} shows the maximal correlation at given $T$ from the search region shown above as a function of $T$, for movies of different blurring. The absolute maxima (red stars in Fig.~\ref{ident_ex}) are also plotted by red stars. A characteristic feature for blurred movies is evident in Fig.~\ref{max_corr_ex}: blurring tends to shift time lags of maximal lensing correlations towards lower values. This does not happen for every fixed point, but it often occurs (see further). In addition, it can be seen that correlations in movies with larger blurring kernels typically reach higher values.
The methodology for automatically identifying lensing correlation peaks, described in this section, will be important mostly in Sec.~\ref{sec_stat}, but we also apply it as a supplementary tool in other sections. 
We hope to further elaborate and refine it in the future, especially when dealing with real data, in order to optimize the detection of correlation enhancements.

\section{What is the minimal resolution required to identify the lensing peak?}\label{sec_blur}
In our previous work \cite{bezdekova26}, we have shown that the correlation peak can be identifiable even when the movie is blurred. Our largest blurring kernel was of FWHM=15~$\mu$as. However, we did not discuss up to which FWHM one can blur the movie while still keeping the correlation peak distinguishable. 

\begin{figure}[h!t]
\centering
\includegraphics[width=.95\textwidth]{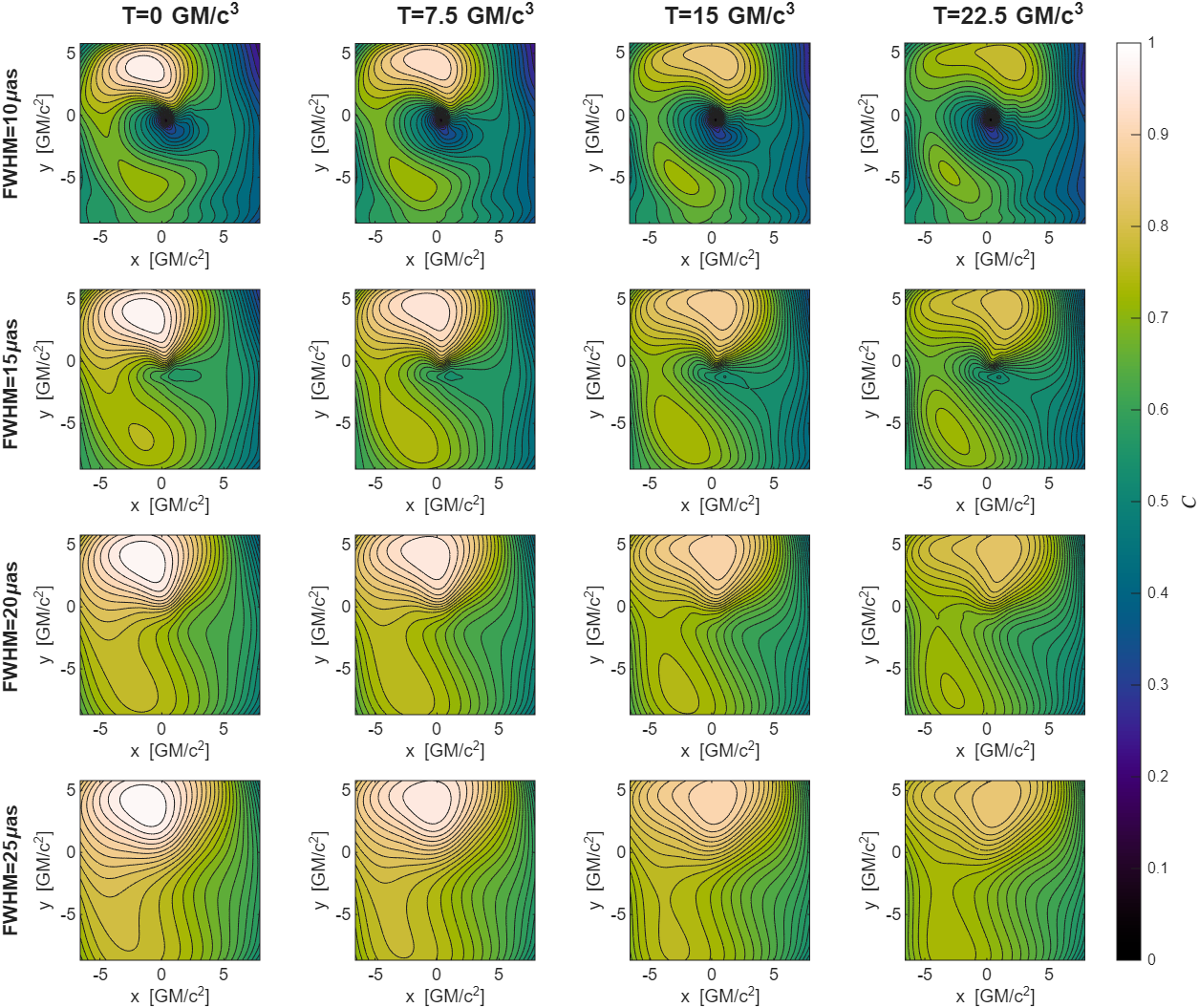}
\caption{Contour plots of correlation maps in the $(x,y)$-plane at the time lags $T=0, \, 7.5, \, 15$ and $22.5~GM/c^3$ (columns) for blurred movies with kernels of FWHM$=10, \, 15, \, 20$, and $25~\mu$as (rows) for the fixed point located at $(x_0',y_0')=(-1.317,3.819)~GM/c^2$.}\label{blur_ex}
\end{figure}

A natural criterion for distinguishability is whether there exists a clear separation between the astrophysical and correlation peaks. 
In principle, a measure of the separation between two peaks in the configuration space may be given a precise mathematical definition. For example, it could be quantified by maximizing, over all possible curves connecting the two peaks, the minimal value of correlation attained along the curve, and then judiciously comparing this minimal value with the values at the peaks. Although such a precise analysis is outside the scope of this paper, here we qualitatively investigate peak separation as a function of blurring. 
Examples of contour plots of the correlation function in the $(x,y)$-plane are shown in Fig.~\ref{blur_ex} for blurred movies with kernels of FWHM=10, 15, 20, and 25$~\mu$as, for time lags $T=0$, 7.5, 15, and 22.5~$GM/c^3$. 
As discussed in \cite{bezdekova26} for the same BH movie, both the astrophysical and lensing peaks are rather elongated in the $T$ direction. 
Nevertheless, their separation in the $(x,y)$-plane often allows us to distinguish the two peaks, provided the blurring is not too large, as was demonstrated in \cite{bezdekova26}.

Here we are interested in investigating up to which level of blurring the peaks are still distinguishable. As Fig.~\ref{blur_ex} demonstrates, the saddle (col) connecting the two peaks at fixed $T$ becomes shallower with larger blurrings.
For example, its depth can be measured by the difference between the values of $\mathcal{C}$ at the lensing peak and the saddle.
It is seen that, roughly speaking, at a blurring with FWHM=
25~$\mu$as the correlation peak becomes indistinguishable from the astrophysical one, hindering the identification of the correlation peak. The precise value of blurring where the peaks become indistinguishable depends on definitions and
varies for a different fixed point. Nevertheless, blurring with FWHM=20~$\mu$as seems to serve as a good threshold for most points, at least in the movie we studied. Since in this dataset, the diameter of the circle-like brightness enhancement is of about 40~$\mu$as ($\sim10~GM/c^2$), it seems reasonable that a movie blurred with FWHM larger than half of this diameter cannot preserve the information needed to identify the lensing peak.

To conclude the discussion, we compare these results with the expected angular resolution of current and upcoming observations. 
The current resolution of EHT is estimated at $\sim20-25~\mu$as \citep[cf.][]{EHT19I,EHT22I}. As for upcoming imaging campaigns, the expected resolution of ngEHT is $\sim15~\mu$as \citep{johnson23}, and its space extensions are expected to provide a significantly higher resolution. For example, BHEX's (nominal) resolution is expected at $\sim6~\mu$as \citep[]{sridharan24}. These values seem to be quite promising for our suggested technique.
Therefore, raw angular resolution 
does not seem to be the main obstacle for our method---at least when one assumes the BH movie is not corrupted in other ways. However, this does not seem to be a good assumption. There are other, more serious limitations that must be taken into account. We discuss some of them in the following sections.

\section{What is the minimal movie duration required to identify the lensing peak?}\label{sec_dur}
The total duration of the BH movie we are using is 4000~$GM/c^3$. 
For M87* this would correspond to $\sim4$~years, while for Sgr A* this is about 22~hours. The long characteristic timescale of very heavy sources such as M87* motivates us to investigate whether it would also be possible to reliably detect the correlation peaks in a movie of shorter duration.

\begin{figure}[h!t]
\centering
\includegraphics[width=.9\textwidth]{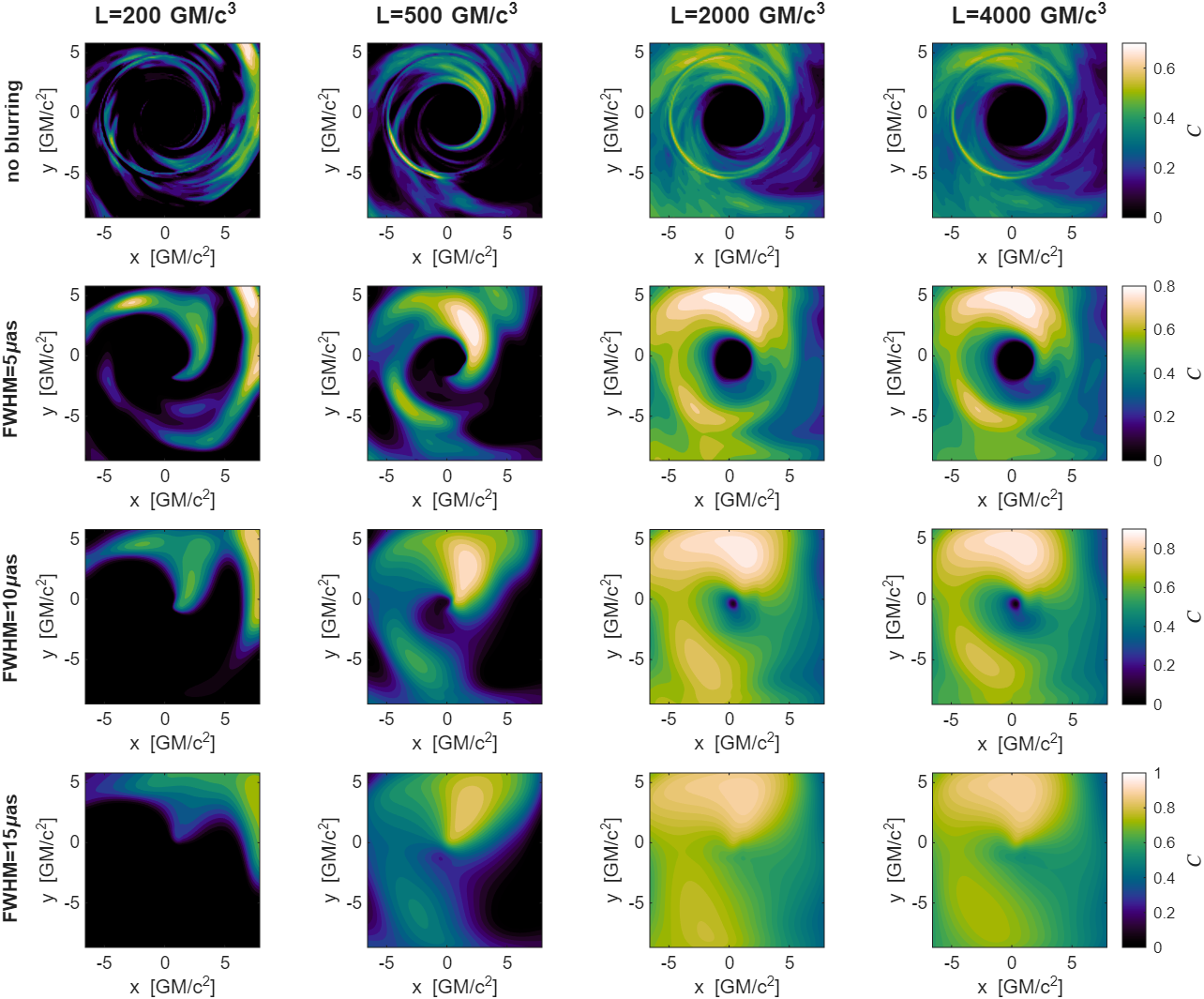}
\caption{Correlation maps at fixed $T=15~GM/c^3$ for different blurrings (rows) and movie durations (columns) for the fixed point $(x_0',y_0')=(-1.317,3.819)~GM/c^2$. As the duration increases (from left to right), stochastic fluctuations become progressively suppressed and the maps stabilize at their large-duration limit, showing the persistent (astrophysical+lensing) correlations in the movie.}\label{dur}
\end{figure}

Fig.~\ref{dur} shows fixed-lag correlation maps, taken at $T=15~GM/c^3$, obtained from segments of our original movie of various durations $L$. In addition to the original unblurred movie, we studied blurred movies with FWHM=5, 10, and 15~$\mu$as. 
For short durations, as is most clearly seen in the leftmost column corresponding to a duration of 200~$GM/c^3$, the correlation maps differ significantly from their long-duration counterparts. This may be understood as a result of the stochastic nature of the flow, which leads to stochastic fluctuations (across the ensemble of flows with the same parameters) in the short-duration maps. As the duration increases, the persistent correlation features (both universal/lensing and astrophysical) progressively dominate the correlation fluctuations, until at some point they become negligible, leaving only the persistent features.
In particular, since lensing features are swamped by fluctuations for short movie durations, the photon-ring-like correlation enhancement along an approximate circle of diameter $\sim10~GM/c^2$ is indistinguishable in their correlation maps. In contrast, in the maps produced from longer-duration movies, this feature is clearly visible.

\begin{figure}[h!t]
\centering
\includegraphics[width=.8\textwidth]{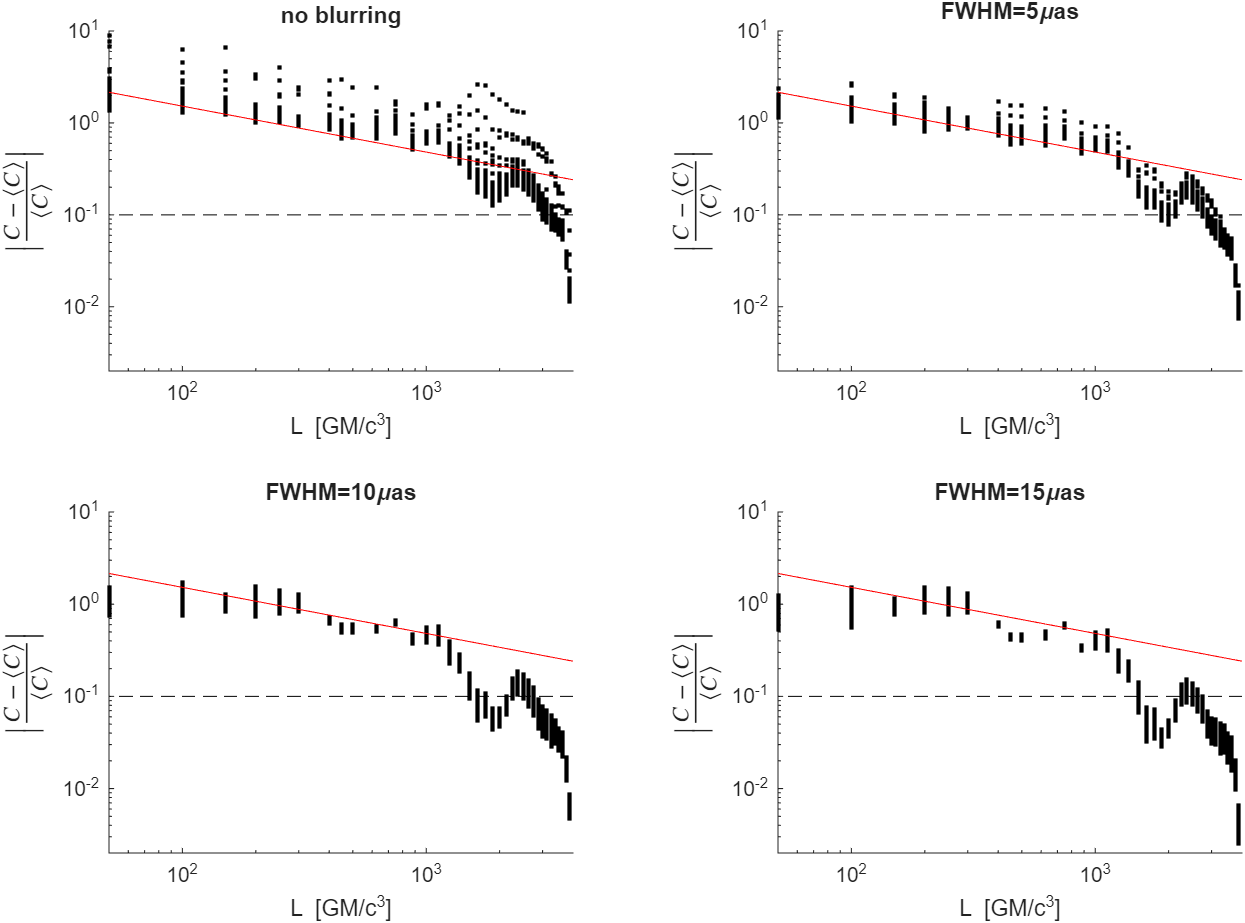}
\caption{Absolute mean normalized difference between expected and calculated correlations for the fixed point $(x_0',y_0')=(-1.317,3.819)~GM/c^2$ at different $T$ as a function of the movie duration $L$. The dashed lines show the value 0.1. The red solid lines are $\propto L^{-1/2}$.
}\label{dur_point}
\end{figure}

In order to quantify the rate of convergence of the correlation maps toward their large-duration limit, in Fig.~\ref{dur_point} 
we display the absolute mean difference (we take the early point to be our fixed point $(x_0',y_0')$ and average over all later points $(x,y)$ that yield positive correlation with the fixed point at $T=0~GM/c^3$) of the correlation function $\mathcal{C}$, calculated using a movie of duration $L$, and the correlations obtained using the full movie duration, which we will refer to as the ``expected value" $\langle\mathcal{C}\rangle$, normalized by $\langle\mathcal{C}\rangle$, for different values of $L$ (horizontal axis). For each duration, we plot different points, corresponding to 81 different lags: $T=0~GM/c^3$ through $T=40~GM/c^3$ in increments of $T=0.5~GM/c^3$. We label all points with the same (black) color to avoid clutter; our goal is to show the range of values assumed by the absolute mean correlation difference.

The results in Fig.~\ref{dur_point} are shown for correlations derived from movies without blurring, and movies blurred with FWHM=5, 10, and 15~$\mu$as (different panels). 
Due to the emission's stochastic nature, we expect the correlation difference to fall off 
$\sim \sqrt{\mathcal{K}/L}$ at large $L$, where $\mathcal{K}$ is a constant characteristic timescale. 
In order to independently determine $\mathcal{K}$, we searched for a temporal correlation length scale in our data set. We fixed it as the mean FWHM of the autocorrelation function of more than 2,500 data points in the original (unblurred) movie, equally distributed on the observer's screen, avoiding the inside of 
the dark region in the center, also known as the ``inner shadow'' \citep{Chael2021}.
This calculation yielded $\mathcal{K}\approx 15.23~GM/c^3$. The function $\sqrt{\mathcal{K}/L}$ is plotted in red in all panels of Fig.~\ref{dur_point}. 
A similar calculation of $\mathcal{K}$ in the blurred movies yields increasingly higher values with increasing blurring. While the red lines in Fig.~\ref{dur_point} clearly demonstrate the anticipated duration dependence $\propto L^{-1/2}$ for short enough durations, as $L$ becomes of the order of magnitude of the full movie duration $4000~GM/c^3$ (very roughly, around $1000~GM/c^3$), the absolute mean normalized difference begins to fall off more rapidly with $L$. This may be understood as a result of the finite size of our full movie, $4000~GM/c^3$. When the duration reaches this maximal value, the absolute mean normalized difference must exactly vanish, by definition; therefore, above some duration, it must decrease more rapidly than $\propto L^{-1/2}$.
Fig.~\ref{dur_point} also displays how, due to the stochastic nature of the movie, the absolute mean normalized difference fluctuates around the $\sim \sqrt{\mathcal{K}/L}$ behavior, with the fluctuations also decreasing as $L$ grows.
Another noticeable feature is the bump around $2500~GM/c^3$. This unexpected increase of correlation differences is caused by a flare present in the movie around this time, which causes non-negligible contribution to correlations. At larger durations after the bump, the correlation difference continues to fall off more steeply.

In the profile for the unblurred movie (top left panel), there are two lag values for which the correlation differences are still significantly larger than $\sqrt{\mathcal{K}/L}$ even for longer movie durations. These correspond to $T=0$ and $38~GM/c^3$. 
As far as we can tell, this seems to be a statistical fluctuation of correlation maps, arising due to the stochasticity of the accretion flow. It occurs only for this pair of lags---a small portion of the lags we studied---and is suppressed for all blurred movies.
Even if such fluctuations are present in a similar analysis performed on a different GRMHD-simulated movie, it seems unlikely that they will arise at the same lag values as here. We leave further discussion of this for future work.
It is also seen that in the blurred cases the maximal value (maximizing over lag at fixed duration) of the absolute mean normalized correlation difference is generally smaller, and in particular it makes its first drop below 10\% (dashed lines) for shorter movies compared to the unblurred case. The larger the blurring, the less scattered are the correlation differences. However, it is necessary to say that the expected values of correlations, $\langle\mathcal{C}\rangle$
are different for different blurrings.

Even though the minimal movie duration suitable for our analysis can be shorter as the blurring increases, the information contained in such a movie is still suppressed because its resolution decreases. 
In other words, the blurring makes the correlation profile smoother and more stable, but hinders the identification of the exact position of maximal correlation at the spatial level. Very roughly, for the present case study, 
it seems quite reasonable to expect that with a movie duration of $2000~GM/c^3$ one can reliably detect the lensing correlation peak; this threshold duration varies to a certain extent with the choice of blurring and fixed point. It would be interesting to study how these conclusions depend on the parameters of the BH and accretion flow.

A movie duration of 2000~$GM/c^3$ corresponds to $\sim 2~$years for M87* and $\sim 11~$h for Sgr A*. This estimate assumes for simplicity that the signal is taken continuously for the whole period which is not feasible, especially in the M87* case. We will discuss in more detail the issue of temporal discontinuities in the measurements in Sec.~\ref{sec_seg}.
For M87*, therefore, the required movie duration for detecting extreme lensing with correlations introduces the challenge of long-term continuous monitoring of the source. The estimated required monitoring time stated above, $\sim 2~$years, does not seem impossible but is an important factor to consider, and estimate more carefully across a wider range of source parameters.
For Sgr A*, the required length of the movie does not seem to be an impediment. 
It is important to note that, if a bright enough localized emission event (flare) occurs close to the BH, in principle it could be possible to extract extreme-lensing information in significantly shorter durations than those quoted above; see, e.g., \cite{tiede20}.

\section{How do the correlation maps depend on movie cadence?}\label{sec:cadence} 
The cadence of our original movie is $0.5~GM/c^3$, which is approximately equal to 4.5~h for M87* and 10~s for Sgr A*. For Sgr A*, this cadence is too fine to be resolved by current EHT or ngEHT. There are two reasons for this. First, EHT measurements are typically collected for several minutes in limited intervals---segments---which themselves are separated by lags of a few minutes. This aspect is addressed in the next section. Second, in order to suppress signal noise, typically an average of the visibility over these segments is performed.
However, when calculating correlations, performing first a time average of the intensity/visibility is not desired, as it necessarily leads to loss of information on real signal fluctuations. Moreover, the time average performed as an intrinsic part of the correlation computation (represented by $\langle \rangle$ in Eq.~\eqref{corr_defT}) can in itself suppress the noise, see Sec.~\ref{sec_noise}.
It therefore seems that reaching the finest possible cadence is preferred over time averaging at the intensity/visibility level. In this section, in order to further test the feasibility of realistic correlation measurements, we will vary the cadence and study up to which value one can still identify the correlation peak.

To set the new cadence, $\Delta t$, we average over $m$ movie snapshots (frames), where $m=\Delta t/(0.5~GM/c^3)$. This procedure has a similar effect to the EHT averaging mentioned above, which mainly aims to reduce the noise level of the obtained data. In our case, at the moment there is no noise, and hence the effect of the averaging is to coarse grain the data in the temporal direction, suppressing its rapidly variable component. 
A second possible way to produce a lower-cadence movie could have been to 
This strategy resembles the segmentation phenomenon, mentioned above,
which is discussed in detail in the next section.

\begin{figure}[h!t]
\centering
\includegraphics[width=.75\textwidth]{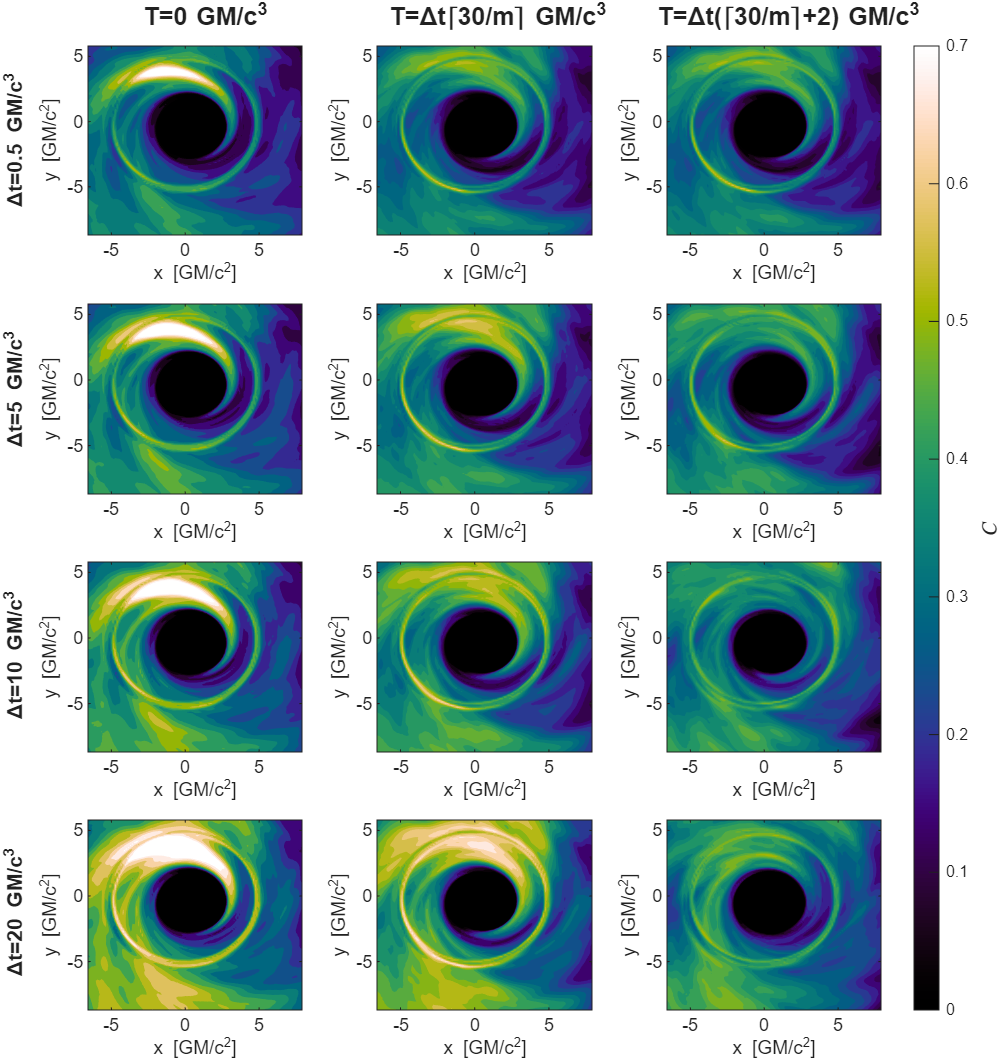}
\caption{Fixed-lag correlation maps computed from unblurred movies with cadence $\Delta t=0.5~GM/c^3 ~(m=1)$, $5~GM/c^3 ~(m=10)$, $10~GM/c^3 ~(m=20)$, $20~GM/c^3 ~(m=40)$ (rows) at the lag values $T$=0, $\Delta t\lceil30/m\rceil$, and $\Delta t(\lceil30/m\rceil$ + 2)~$GM/c^3$ (columns), where $\lceil\,\rceil$ returns the smallest integer greater than or equal to a given real number.}\label{av_corr_dt}
\end{figure}

We have calculated the correlation function \eqref{corr_defT} for the fixed point $(x_0',y_0')=(-1.317,3.819)~GM/c^2$ and several values of movie cadence $\Delta t$ 
in order to study how the cadence affects the ability to identify the correlation peaks. In Fig.~\ref{av_corr_dt}, a few examples of fixed-lag correlation maps 
are shown, for the cadence choices: $\Delta t=0.5~GM/c^3 ~(m=1)$, $5~GM/c^3 ~(m=10)$, $10~GM/c^3 ~(m=20)$, $20~GM/c^3 ~(m=40)$. The correlation maps are shown at $T=0$, $\Delta t\lceil30/m\rceil$, and $\Delta t(\lceil30/m\rceil$ + 2)~$GM/c^3$, where $\lceil\,\rceil$ returns the smallest integer greater than or equal to a given real number. Note that the middle value was set such that for $\Delta t=0.5~GM/c^3$ it corresponds to 15~$GM/c^3$. 
Our results show that the spatial correlation structure is quite stable to change of cadence, across a wide range of values. This remains true even though in the unblurred maps a mild, roughly overall increase in correlation is seen when the cadence is decreased (equivalently, $\Delta t$ increased). In the blurred maps, to be discussed below, this effect is much weaker, up to a degree that it is barely noticeable.

Despite the fact that the correlation structure is maintained even at low cadence, one crucial aspect is lost when increasing $\Delta t$. Since the minimal nonzero step size (in absolute value) on the lag axis is determined by the cadence, as the cadence becomes low (values of $\Delta t$ become large), the temporal resolution on the lag axis decreases, making it difficult to resolve the lag dependence of correlation features.
Moreover, since the lensing peak can get quite close to the astrophysical peak, these two effects might mix (see the last row of Fig.~\ref{av_corr_dt}) and it becomes more challenging to distinguish them from each other.

\begin{figure}[h!t]
\centering
\includegraphics[width=.75\textwidth]{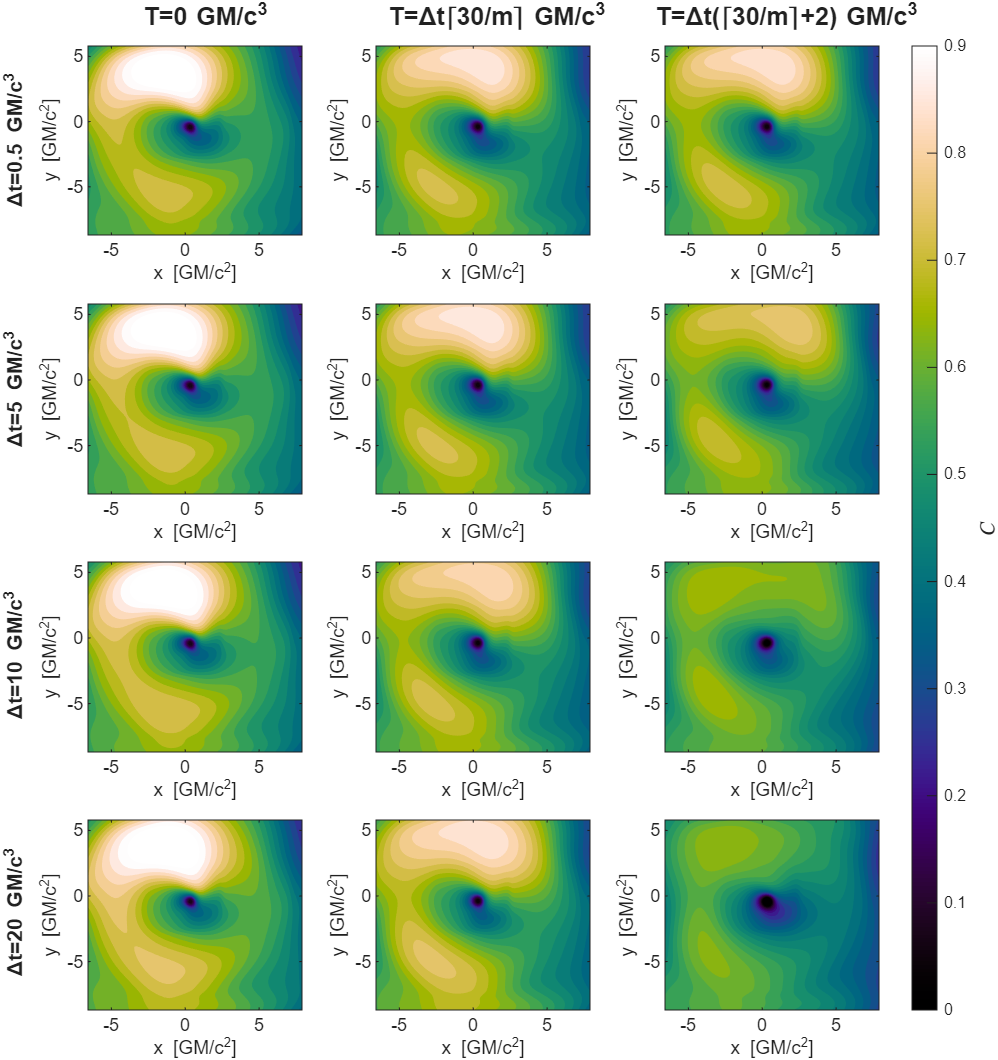}
\caption{Fixed-lag correlation maps computed from blurred movies with FWHM=10~$\mu$as. The choices of values for cadence (rows) and lags (columns) are the same as in Fig.~\ref{av_corr_dt}.}\label{av_corr_b}
\end{figure}

An example of the effect of cadence reduction on correlation maps of blurred movies, here with FWHM=10~$\mu$as, is shown in Fig.~\ref{av_corr_b}. 
Due to the spatial smearing of correlations because of the blurring, the difference between the 
maps due to change of cadence is not as apparent as in the unblurred case. 
For this example, an identification of the lensing peak seems to be possible even in the lower-cadence cases.
Establishing the exact time lag of maximal lensing correlation naturally remains challenging. For different choices of fixed point $(x_0',y_0')$, the results seem to be consistent with the findings presented in this section; we have checked several different examples.

The results of this section provide evidence that probing the extreme-lensing properties of Sgr A* using correlations could be possible even with a cadence as low as $\sim 1$~min. This sharpens the order-of-magnitude estimate for the cadence required to perform such an observation with upcoming missions. Cadence is not expected to be a significant impediment for a similar observation of M87*.

\section{What is the effect of temporal segmentation?}\label{sec_seg}
One of the crucial aspects of current EHT measurements is their temporal segmentation \citep[e.g.,][]{eht19iii}. Due to the rotation of Earth, it is necessary to fine-tune the instrument once in a while to guarantee it is properly calibrated. This leads to gaps (missing data points) in the data time series, which occur each $\sim$ several minutes and also last $\sim$ few minutes. This procedure is unavoidable and it needs to be taken into account when making our data set closer to realistic observations.

To demonstrate how segmentation affects correlation maps, we introduce two free parameters: $d_l$, that denotes the duration (length) of each segment where data points are retained, and $d_s$, that denotes the duration of the space between individual segments. In our calculations, data points within the space of length $d_s$ were substituted by NaNs. The correlation function \eqref{corr_def_discr} was then computed for this redefined array. In real situations, the durations of $d_l$ and $d_s$ might slightly change for each sequence. Nevertheless, in order to keep the model simple, we kept $d_l$ and $d_s$ fixed over the whole data set.

\begin{figure}[h!t]
\centering
\includegraphics[width=.495\textwidth]{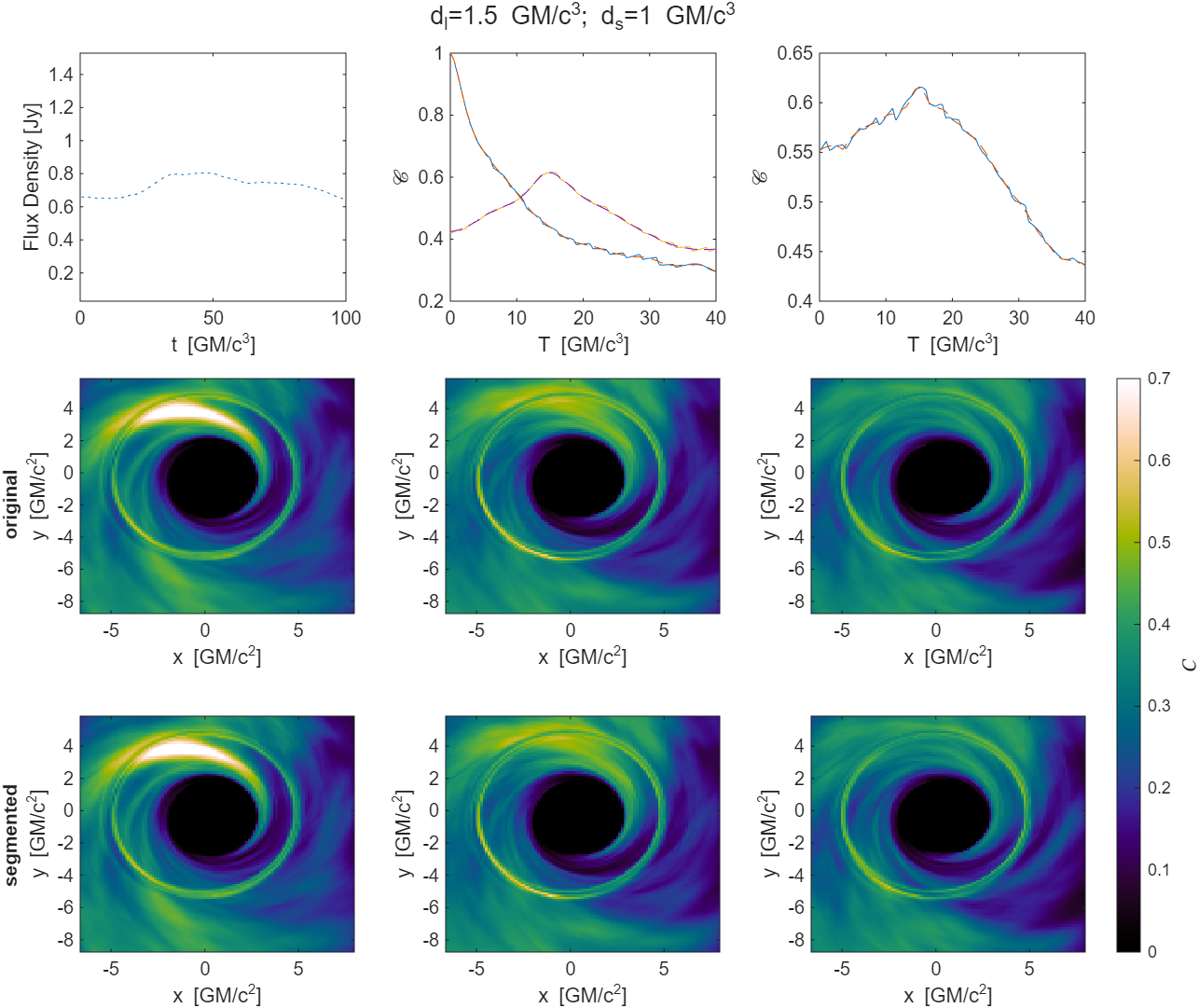}
\includegraphics[width=.495\textwidth]{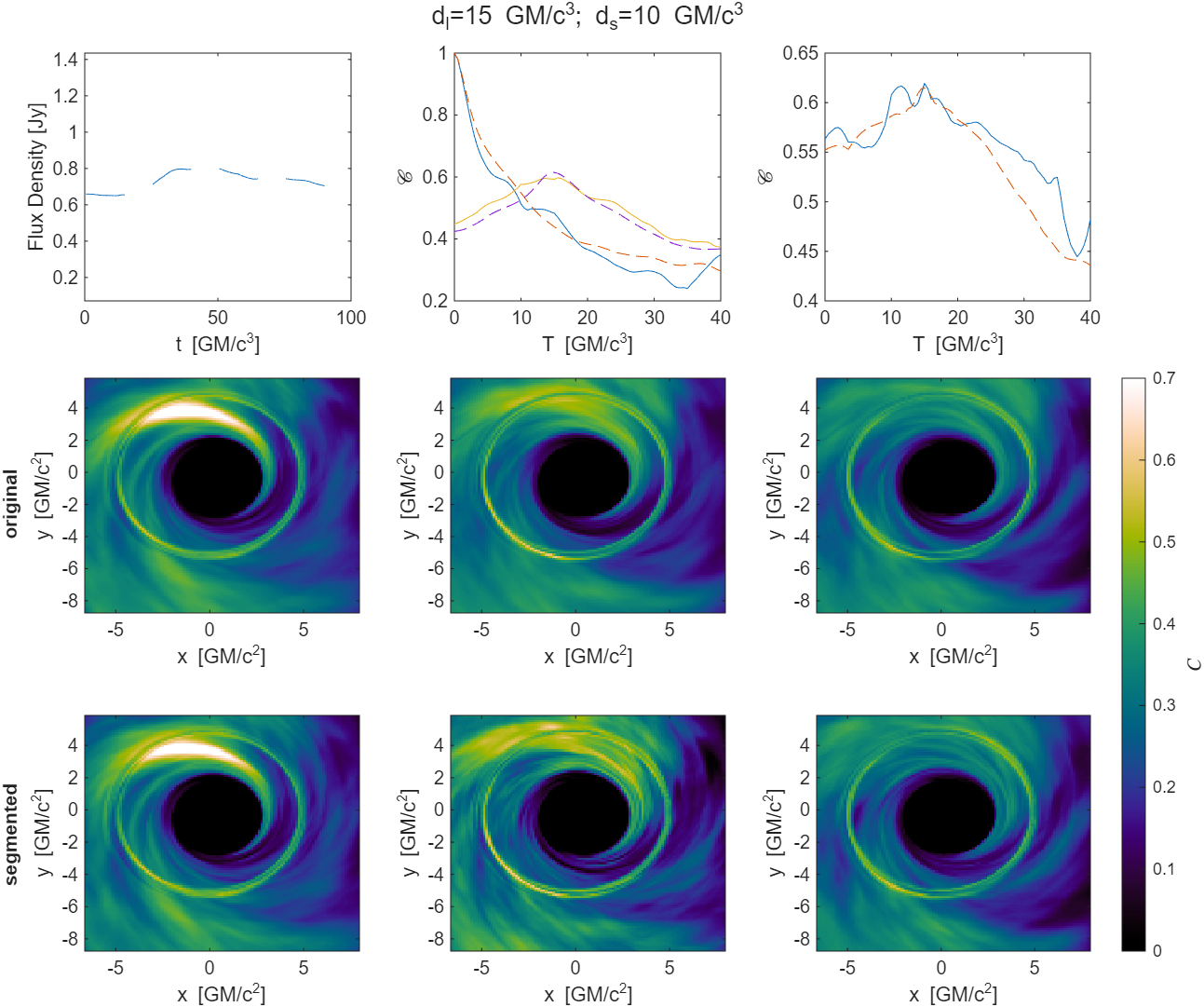}
\caption{The effect of segmentation on correlation maps of the unblurred movie, for two combinations of the parameters $d_l$ (duration of each segment), and $d_s$ (separation between segments). Top left panels: a $100~GM/c^3$ interval of the light curves after the segmentation. Top middle panels: autocorrelation at the fixed point, $\mathcal{C}(T,x_0',y_0',x_0',y_0')$ (dashed red: unsegmented, solid blue: segmented) and cross-correlation with the lensing point $\mathcal{C}(T,x_\ell,y_\ell,x_0',y_0')$, where the lensing point $(x_\ell,y_\ell)$ is defined as the screen position at which the $n=1$ image would appear for a pointlike equatorial source that has an $n=0$ image at the fixed point $(x'_0,y'_0)$ (dashed purple: unsegmented, solid yellow: segmented). Top right panels: maximal correlations found across the $(T,x,y)$-space within the lensing region, taken here to be $|\delta \varphi|<\pi/3$ (see Sec.~\ref{sec_ident}; dashed red: unsegmented, solid blue: segmented). Middle and bottom rows: correlation maps from the original and segmented movies at $T=0,\,T_{max},\,T_{max}+10~GM/c^3$, where $T=T_{max}$ corresponds to the same lag where $\mathcal{C}$ is maximal in the top right panel.}\label{seg_p1}
\end{figure}

\begin{figure}[h!t]
\centering
\includegraphics[width=.495\textwidth]{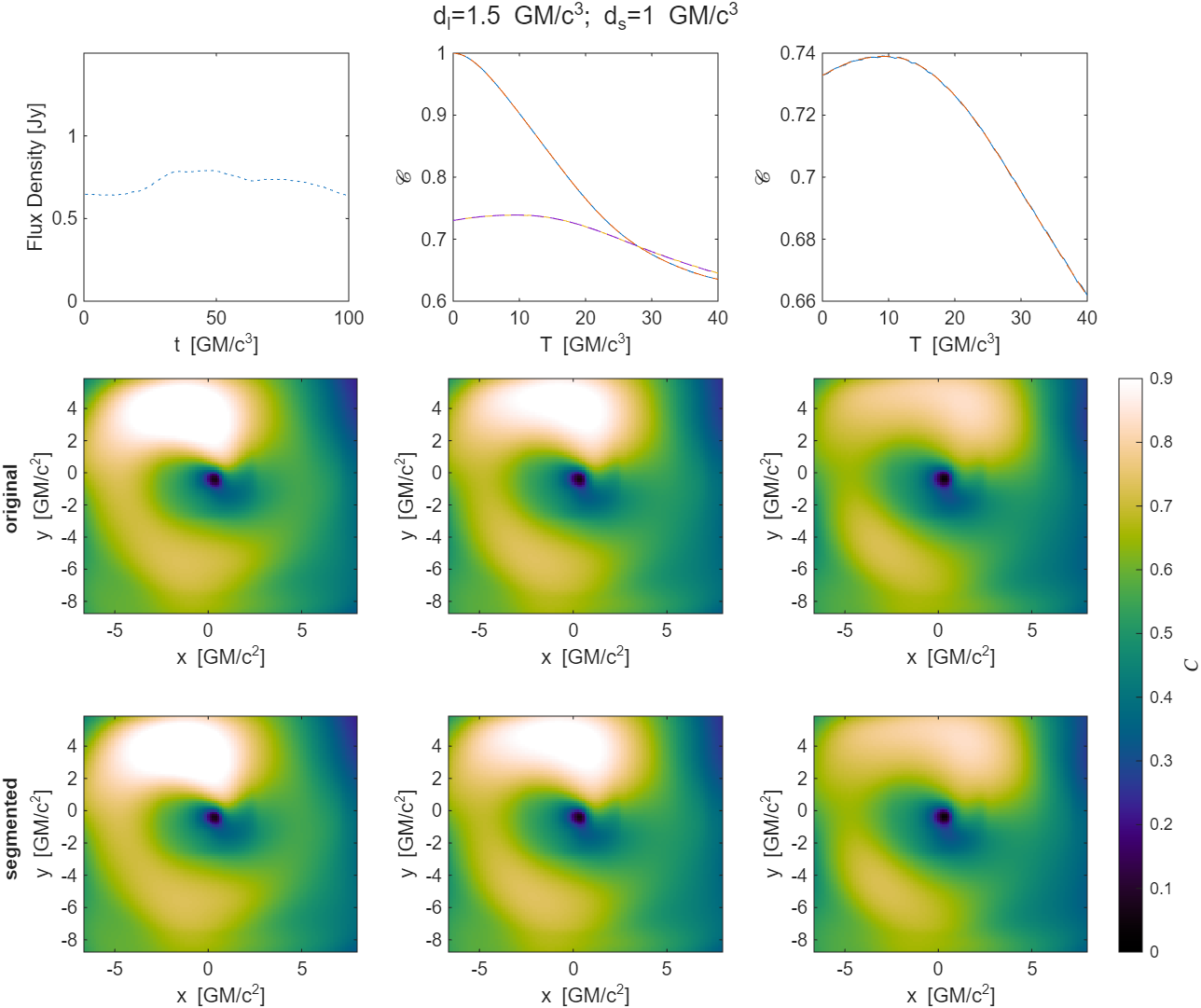}
\includegraphics[width=.495\textwidth]{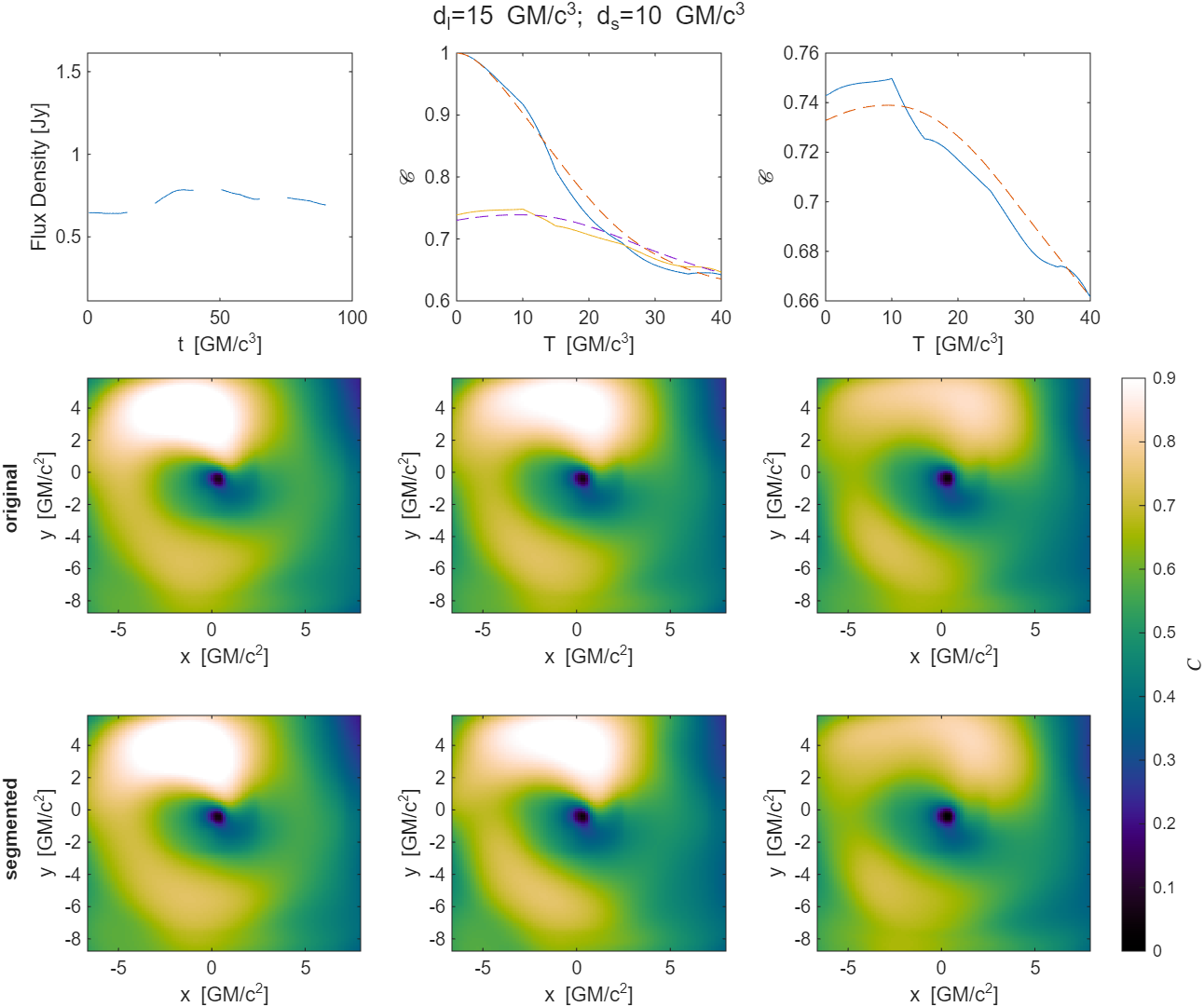}
\caption{The effect of segmentation on correlation maps of blurred movies with FWHM=10~$\mu$as. All conventions are the same as in Fig.~\ref{seg_p1}.
}\label{seg_p2}
\end{figure}

Figs.~\ref{seg_p1}, \ref{seg_p2} show correlation maps calculated for segmented data sets with two choices of $d_l$ and $d_s$ and compare them with results from the original (unsegmented) data set for unblurred and blurred (with FWHM=10~$\mu$as) movies, respectively. 
This comparison demonstrates that, at the level of spatial fixed-lag correlation maps, the general correlation structure is rather stable under segmentation. The only case in which the effect of segmentation is noticeable is in the unblurred case, and when $d_l$ and $d_s$ are of a similar order of magnitude as the lensing timescale (right-hand side of Fig.~\ref{seg_p1}).
Even in that case, on average, the correlations are not significantly decreased due to the segmentation, and the lensing peak remains distinguishable from the astrophysical one. It turns out that in some cases the maximal (normalized) correlation value of the lensing peak is even increased in the segmented movie. However, when analyzing the lag dependence of auto- or cross-correlations between particular pixels in the image (upper middle and upper right panels), the effect of segmentation becomes more apparent.

The problem that arises when looking at such quantities
is the occurrence of false peaks or kinks, for both blurred and unblurred movies. Since the number of data points summed over in the correlation calculation at each lag value changes abruptly as a function of lag due to the segmentation, 
there can occur artificial increases of (normalized) correlations. Such a behavior was already observed in the autocorrelations of measured light curves for Sgr A*, see \cite{wielgus22,cardenas24}. Here we demonstrate once again that this behavior can indeed significantly affect the analysis of correlation functions and introduce misleading features. When designing the observation and/or interpreting the data, one has to be very careful not to interpret these artificial features as genuine correlation peaks.
On the other hand, as the examples in Figs.~\ref{seg_p1}, \ref{seg_p2} demonstrate, and may be expected, if the duration of both segments and spaces is small enough (namely, relative to the lensing timescale), segmentation will not have a significant effect on the correlations

An extreme case, not explicitly shown here, is when the segments are shorter than the spaces between them. This case does not appear to occur in current EHT, but it is worth mentioning since in principle it can cause confusion. In such a situation, there will be time lags at which the correlation cannot be calculated (the normalized correlation will not be defined), just because of the missing points. 
In real data, where the segment durations and the spaces' durations are themselves changing with time, such a situation still might introduce spurious features to the normalized correlation. In addition to the possibility of not being able to compute the correlation function at certain lags, there might be lags in which the number of points summed over in the correlation computation is too small (even if nonzero), making the result at these lags less reliable.

It should be pointed out that the analysis of this section used the full duration of the movie which we have at hand.
If the duration of the movie was shorter, the spurious peaks/kinks arising from segmentation might become more substantial. However, we have verified that when the durations of both segments and gaps remain short enough, the correlation structure remains fairly stable under segmentation.

As mentioned above, segmentation with a timescale (of both segments and spaces) of $\sim$ minutes is a characteristic feature of current EHT data. 
For M87*, this is basically no obstacle, since the timescale is much shorter than the gravitational/lensing timescale.
In the case of Sgr A*, due to the unfortunate similarity of timescales, segmentation must be carefully considered as a potential source of problems. In order to reliably use correlations as a tool for analyzing movies of Sgr A* with upcoming/future interferometric imaging campaigns, it would be desirable to reduce the duration of spaces between segments---namely, by speeding up calibration. 
On the other hand, a sufficiently long movie duration may help confront these obstacles. The other side of the short-timescale coin is that it makes it easier to capture longer movies of Sgr A* (measured in units of the gravitational/lensing timescale).

\section{How does noise injection affect correlation maps?}\label{sec_noise}
In this section, we explore the effects of simulated measurement noise on the correlation function \eqref{corr_defT}. We use several different prescriptions to generate the simulated noise, as discussed below, as well as different values for the parameters that characterize it. We then inject the noise to the signal, $I(t,x,y) \to I(t,x,y) + N(t,x,y)$ and compute $\mathcal{C}$ \eqref{corr_defT}. 
One could equally well define noise in visibility space, i.e., via $V(t,u,v) \to V(t,u,v) + \tilde{N}(t,u,v)$, where $\tilde{N}$ is the 2D spatial Fourier transform of $N$. The properties of the Fourier transform imply, of course, that either definition may be used to describe any noise function. However, some choices of noise functions are more naturally described in one of these bases compared to the other. 
In particular, since VLBI observations naturally measure visibilities, specifying the noise at particular baselines in visibility space seems to be the preferred choice. In order to appropriately model both the signal and noise, it would be desirable to use a finite set of time-dependent visibilities, reflecting the number of baselines expected to be available in a real observation. However, in this work we will not consider finite $(u,v)$-coverage effects and the modeling schemes used to produce the images \citep[e.g.,][]{blackburn20,medeiros23}. 
In the remainder of this section, we study injections of different noise functions which are naturally defined in either the intensity or visibility domains. 
Following {\tt{ngehtsim}} \citep{pesce24}, we mimic thermal noise in each pixel in the image or $(u,v)$-plane, at each time in the movie. The noise is randomly generated from a normal distribution with zero mean and unit variance, returning a random dimensionless number $X$. This number is then multiplied by a constant factor, $\eta$, controlling the amplitude of the noise. In addition, we multiply by other factors of choice: in part of the models, we multiply by the standard deviation of the intensity/visibility of the pixel in the image/$(u,v)$-plane, respectively. In others, we prescribe a particular (linear) scaling with baseline length.
We inspect the correlation maps corresponding to different values of $\eta$ in  given model, and this allows us to estimate the noise levels at which the lensing signatures cease to be identifiable.

\subsection{Intensity Domain} \label{sec:intensity domain}
A natural intensity-domain definition of the noise is $N(t,x,y)= \eta\,X(t,x,y) \, \text{std}\left[I(t,x,y)\right]$, where as explained above, $\eta$ is a dimensionless parameter setting the overall amplitude, $X$ is a random dimensionless number generated at each pixel and time step, and std is the standard deviation---here, of the intensity fluctuation at pixel $(x,y)$. This latter factor enforces the noise amplitude at a given screen location to scale proportionally to the real signal fluctuations at that point. 
\begin{figure}[h!t]
\centering
\includegraphics[width=.495\textwidth]{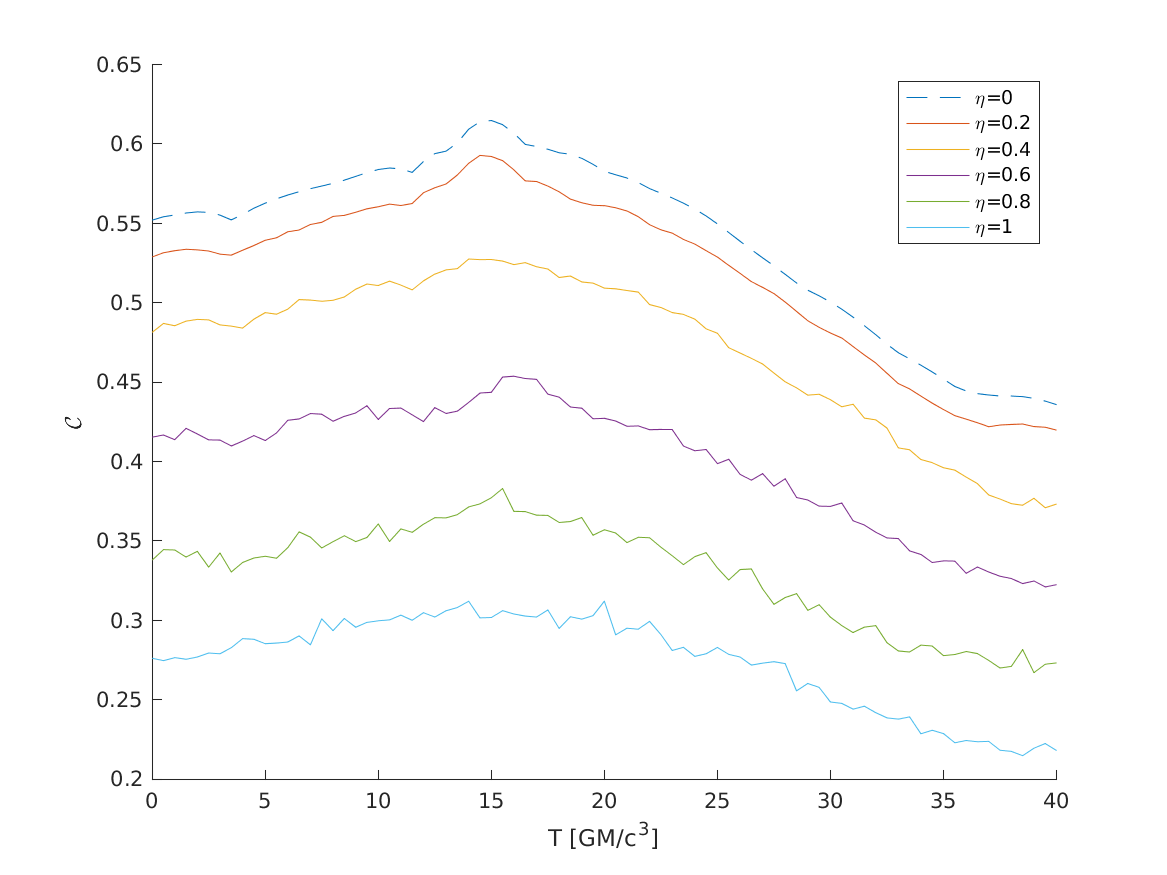}
\includegraphics[width=.495\textwidth]{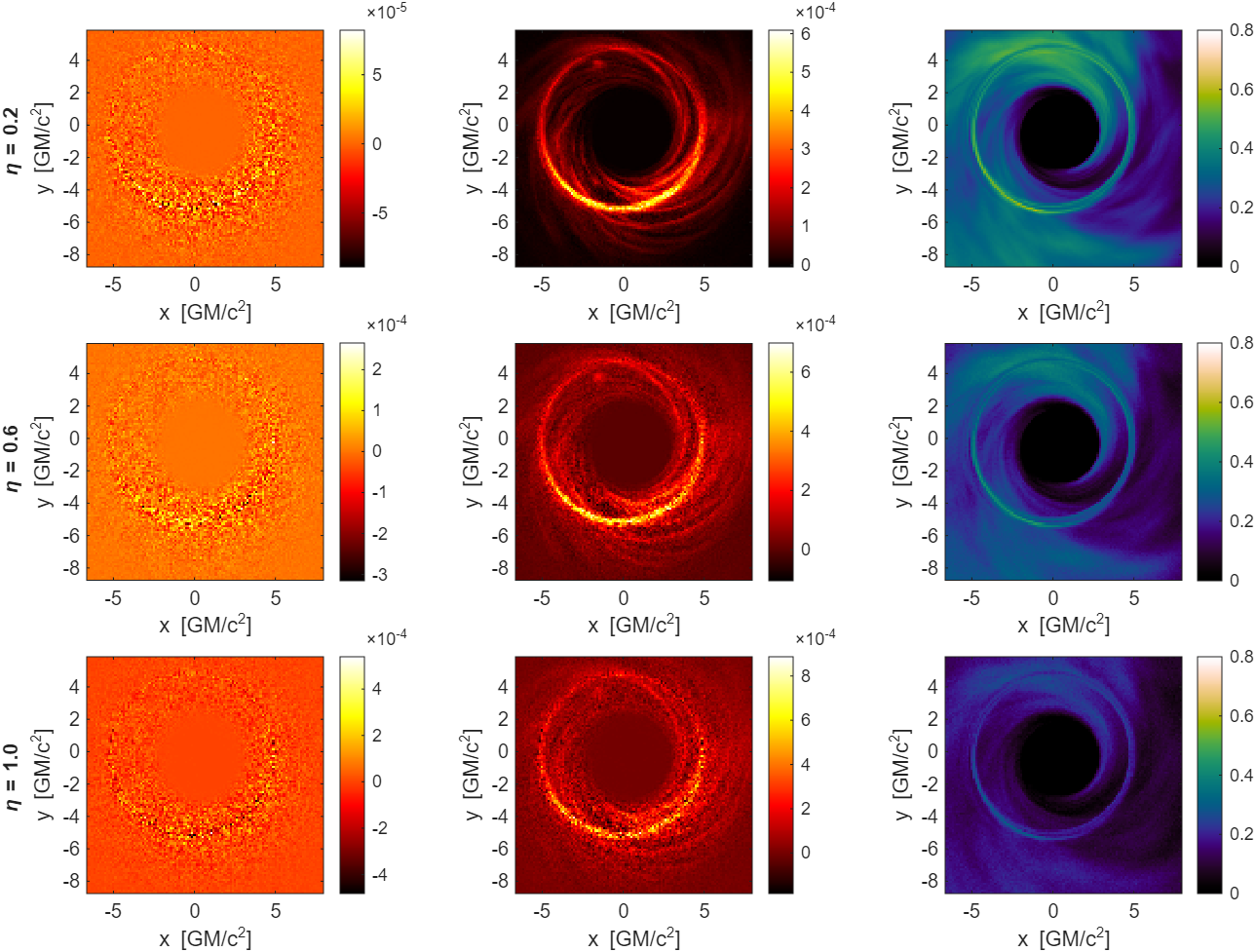}
\caption{(left) Maximal correlation value in the lensing region (part of the screen where lensing signatures appear; see Sec.~\ref{sec_ident}) as a function of time lag, derived from movies with injected intensity domain-motivated noise, defined in Sec.~\ref{sec:intensity domain}, with various noise amplitudes $\eta$. (right) Examples of noise snapshots (left column), the respective noisy movie snapshot (signal + noise, middle column), and the spatial correlation map, taken at the lag $T_{max}$ at which the correlation in the lensing region is maximal (right column).}\label{noi_int}
\end{figure}
The maximal correlation values in the lensing region (the region in which we search for the correlation peak, see Sec.~\ref{sec_ident}) as a function of time lag, in an unblurred movie for different values of $\eta$, are shown in the left panel of Fig.~\ref{noi_int}. It is seen that higher noise levels (larger values of $\eta$) generally lead to a decrease of correlation. Moreover, the maximal value increasingly fluctuates, making it more challenging to reliably identify the lensing peak. On the other hand, in general the noise level does not significantly change the general lag dependence, including the position of the maximum.
In the right-hand side of Fig.~\ref{noi_int}, we present several examples of the noise, with different values of $\eta$. In addition, we show respective (arbitrarily chosen) noisy snapshots along with the resulting correlation maps at the time lags of maximal correlation in the lensing peaks. 
The right column again displays, as expected, a general decrease of correlation with increasing $\eta$. The peaks appear to be increasingly difficult to identify as $\eta$ increases.

\begin{figure}[h!t]
\centering
\includegraphics[width=.495\textwidth]{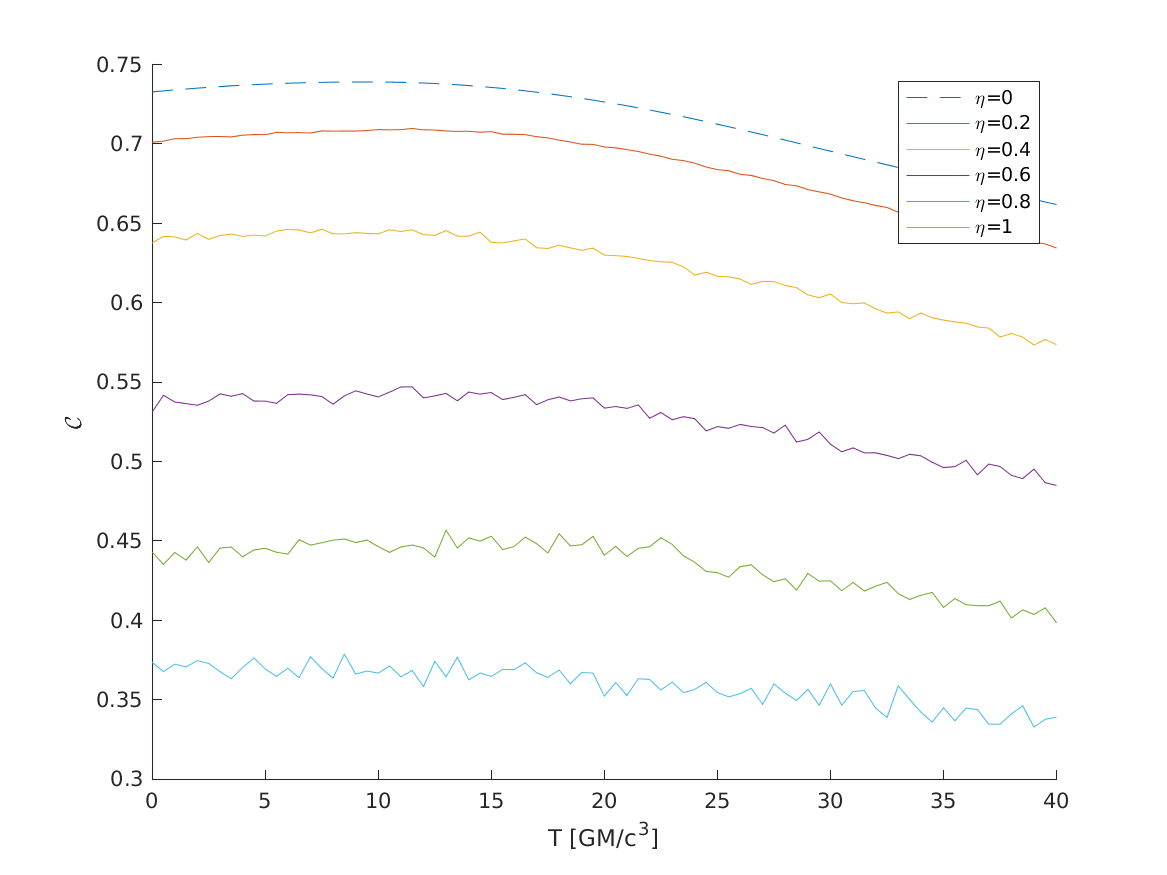}
\includegraphics[width=.495\textwidth]{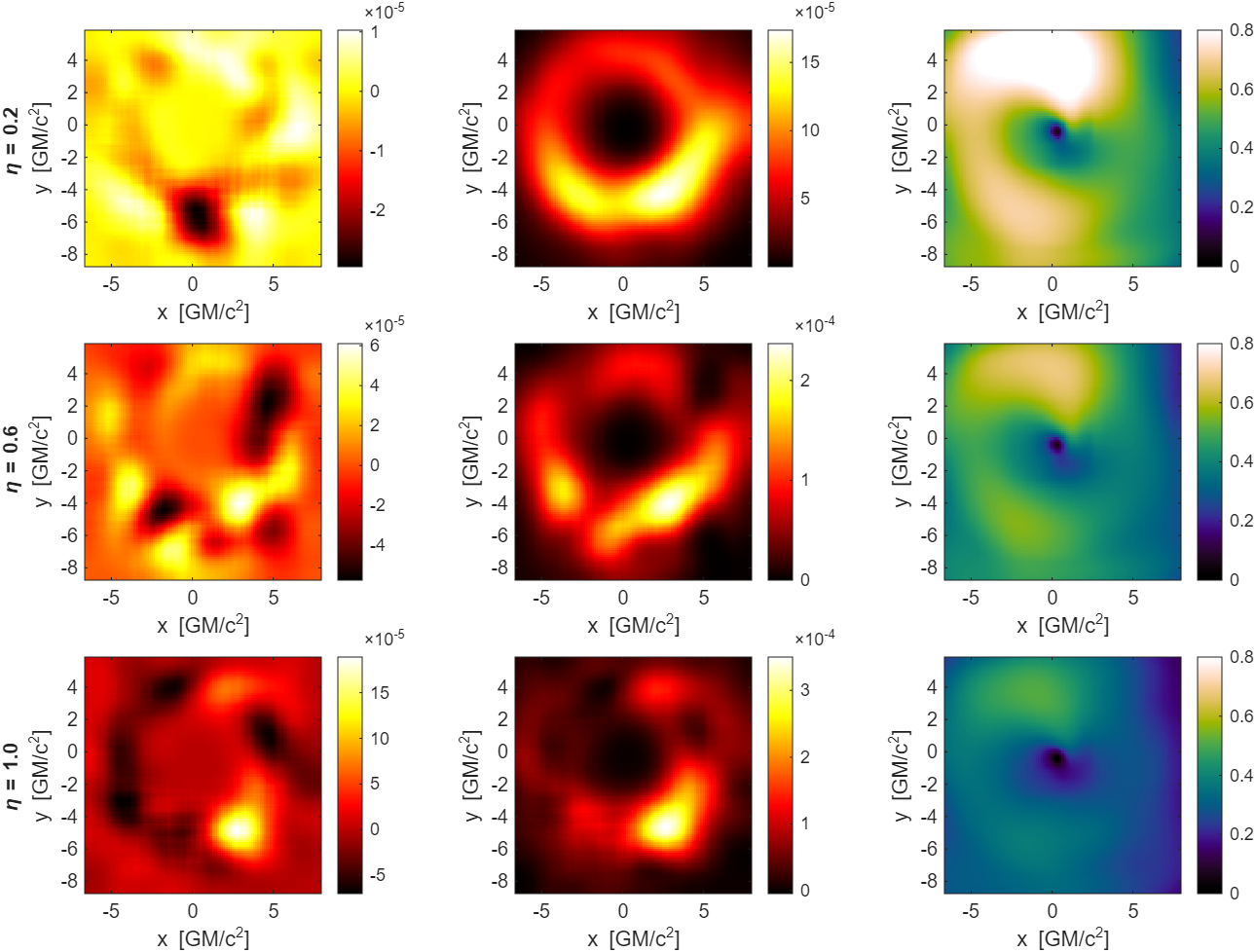}
\caption{Similar plots to those presented in Fig.~\ref{noi_int}, with the movie and noise blurred with a kernel of FWHM=10~$\mu$as.}\label{noi_int_b10}
\end{figure}

Similar plots for the case of blurred noise injected to a blurred movie are shown in Fig.~\ref{noi_int_b10}. Both were blurred with a kernel of FWHM=10~$\mu$as. 
After blurring, the noise was normalized 
to counter a suppression $\sim \left(\mathrm{FWHM}/\mathrm{pixel\, size}\right)^{-1}$ due to the coarse graining associated with blurring.
The correlations are generally higher in blurred movies and the higher-$\eta$ noise 
corrupts the correlations less drastically in comparison to the unblurred case. On the other hand, as is always the case in blurred-movie correlations, 
the peaks are flatter across the $(T,x,y)$-space. 
Since the noise causes random fluctuations of the correlation maps, for large enough $\eta$ at fixed movie duration, false peaks might be identified as lensing peaks. However, for moderate values of $\eta$, the correlation profile and peak can still be roughly identified.

Naturally, the value of the noise amplitude $\eta$ allowing robust detection of a correlation peak (say, for example, a peak with maximal correlation $\mathcal{C}>0.5$)
will depend on the choice of fixed point. If the original maximal lensing correlation was high enough, also $\eta$ can be higher than for originally lower maximal correlations. 
Very roughly speaking, for the movie analyzed here, it appears that even for order-unity noise levels, say, $\eta<0.5$ in the intensity domain, the correlation peaks are still identifiable for most fixed points in our data set, for both the blurred and unblurred cases.

\subsection{Visibility Domain}\label{sec:visibility domain}
As mentioned above, defining the noise in the visibility domain is somewhat more natural since VLBI observations provide visibilities---more precisely, one visibility per baseline.
Here, we do not attempt to mimic the aspect of finite $(u,v)$-coverage, so we do not limit the data to a small number of available visibilities/baselines. Instead, we kept the full available range of visibilities of the original movie, extending to unrealistically large values which will not be detectable in upcoming missions (up to $200~G\lambda$). Hence, this section should be viewed as a test of our techniques, but one that is still not fully realistic. In order to provide a more realistic test, restriction of the $(u,v)$-coverage would be needed. This discussion exceeds the scope of the present paper; we plan to address it in future work.

\begin{figure}[h!t]
\centering
\includegraphics[width=.75\textwidth]{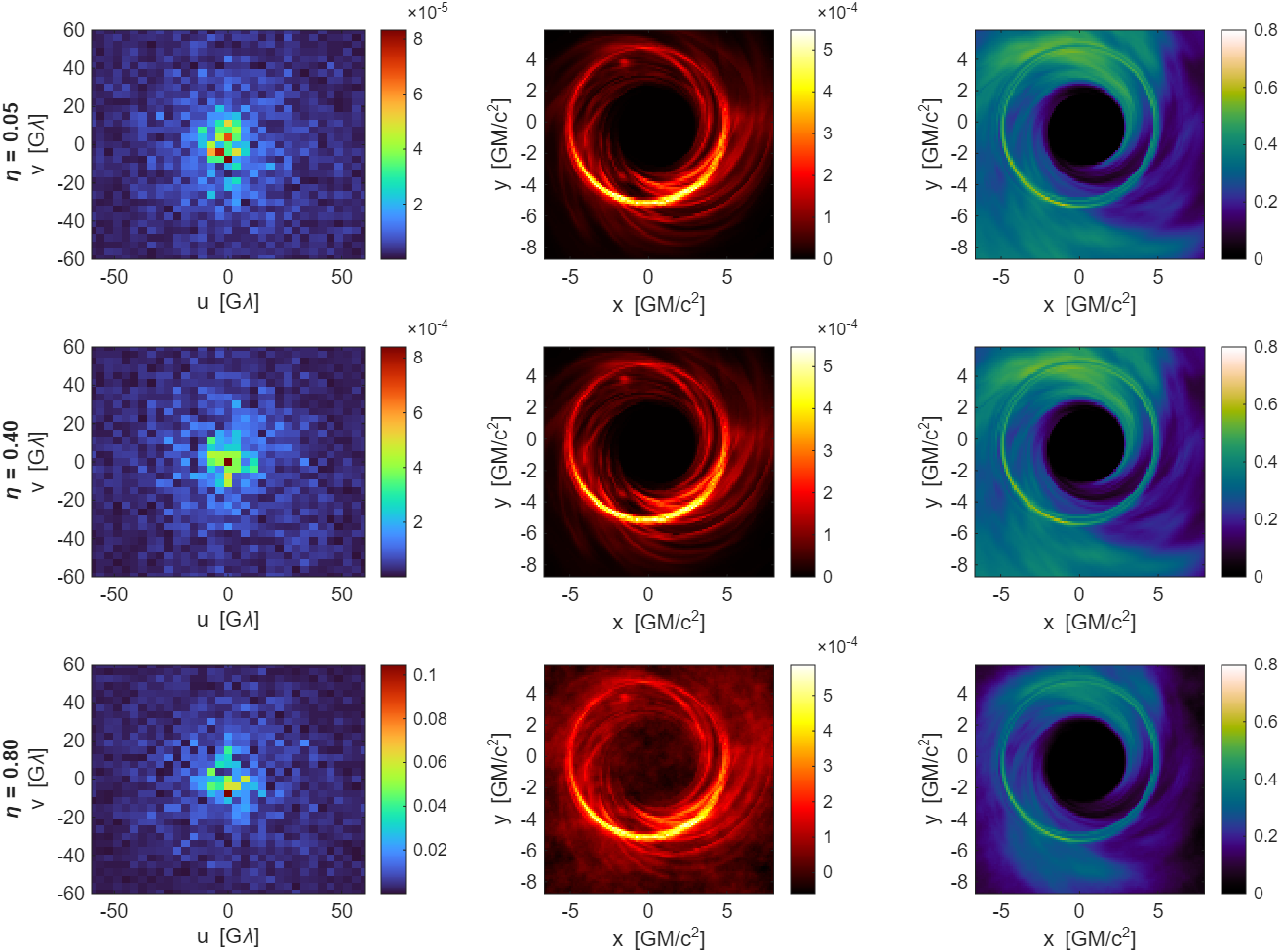}
\caption{Results derived using the noise function $\tilde{N}_1$, defined in Sec.~\ref{sec:visibility domain}. 
Panels show the absolute value of the noise in visibility space, $|\tilde{N}_1(u,v)|$ (left column), the respective noisy snapshots after Fourier transform into the intensity domain (middle column), and correlation maps at the lag $T_{max}$, where the correlation in the lensing region is maximal (right column).} \label{noi_vis_std}
\end{figure}

We study two noise functions which are naturally defined in the visibility domain.
The first is $\tilde{N}_1(t,u,v)= \eta /2 \,\left(X(t,u,v)+i Y(t,u,v)\right) \, \text{std}\left[ \, |V(t,u,v)| \, \right]$, where $X$ and $Y$ are random numbers generated as described above, and here the noise is taken to scale like the standard deviation of the visibility amplitude. The results for correlations with this noise prescription, defined in a similar (yet not identical) way to that described in Sec.~\ref{sec:intensity domain} are shown in Fig.~\ref{noi_vis_std}.

The panels shown in Fig.~\ref{noi_vis_std} are organized in the same way as in the previous subsection, but the noise is now shown in the visibility domain. 
Since the visibility amplitude peaks close to the origin of the $(u,v)$-plane, this noise function is dominated by short-baseline visibilities.
Since the origin of the $(u,v)$-plane (zero baseline) corresponds to the total flux (light curve) 
even though the total light curve of such data is significantly noisy (not shown), the ring's spatial structure in individual snapshots is approximately unaffected by this noise contribution (middle column of Fig.~\ref{noi_vis_std}). Furthermore, as seen in the correlation maps (right column of Fig.~\ref{noi_vis_std}), despite the relatively high noise level (high $\eta$), the correlation structure is still significant and the peaks are identifiable.

\begin{figure}[h!t]
\centering
\includegraphics[width=.75\textwidth]{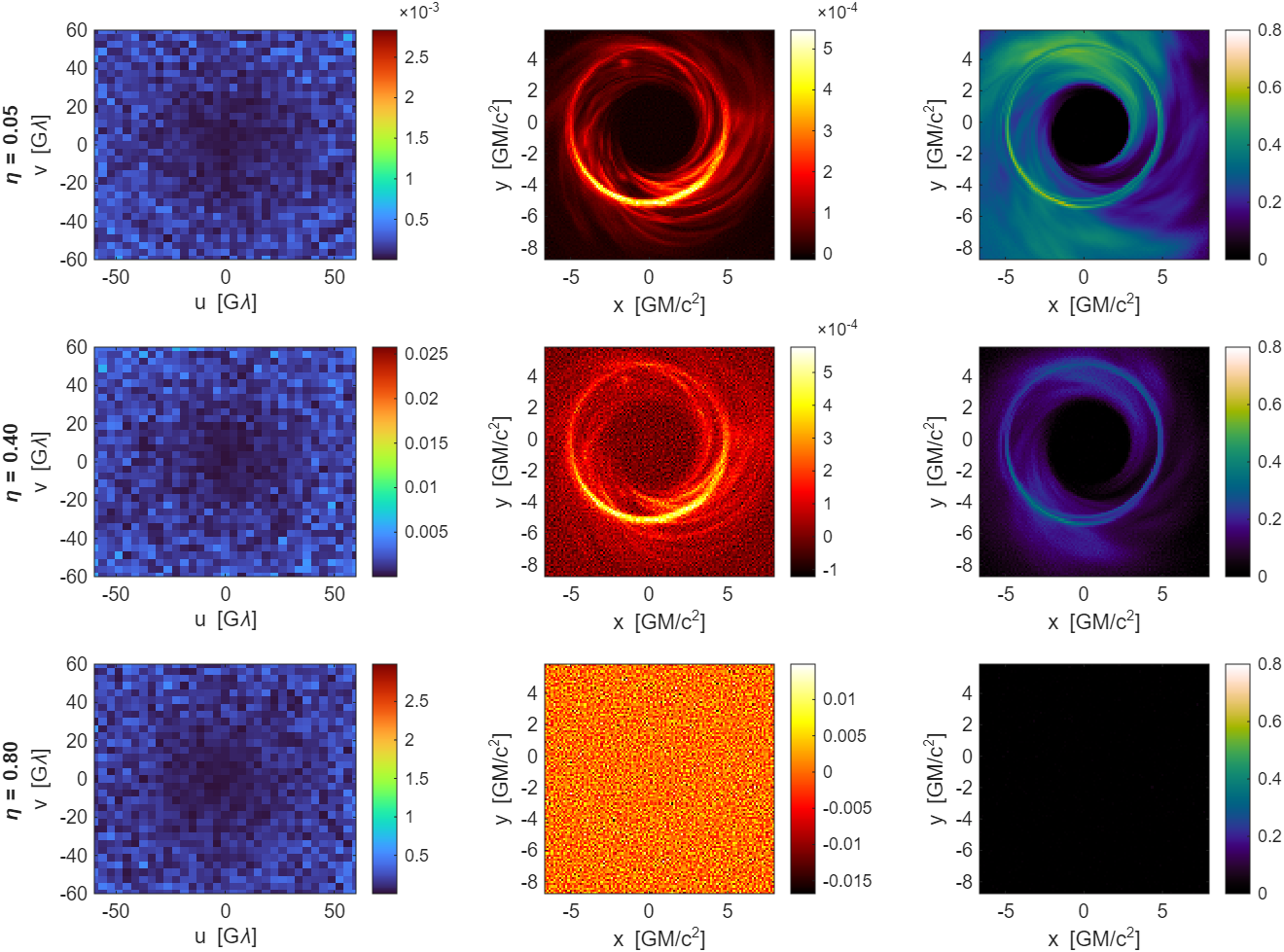}
\caption{Similar plots to those presented in Fig.~\ref{noi_vis_std}, produced for the noise function $\tilde{N}_2$ (defined in Sec.~\ref{sec:visibility domain}) where the noise visibility amplitude scales linearly with baseline length.}\label{noi_vis_lin}
\end{figure}

In the second noise prescription we study in this section, we assumed a linear scaling with baseline length, $\tilde{N}_2(t,u,v)= \eta\,\left(X(t,u,v)+i Y(t,u,v)\right) \left( c_1 \sqrt{u^2+v^2} + c_2 \right)$, where $c_1=0.0025$ and $c_2=0.0147~G\lambda$. This form, including the constants, were motivated by real measurements \citep[e.g.,][]{eht19ii}, but for our purposes it only serves as a toy example. The substantial difference compared to the previous case lies in the fact that the noise contribution is generally larger at longer baselines. The results obtained with this noise prescription are shown in Fig.~\ref{noi_vis_lin}.

There is an important qualitative difference between the results obtained for $\tilde{N}_1$ and $\tilde{N}_2$. While $\tilde{N}_1$, which is prominent at short baselines, affects only weakly the lensing correlation signatures, $\tilde{N}_2$ is prominent at long baselines; it tends to disrupt the lensing correlation signatures even at lower amplitude. This is apparent mainly in the middle row of Fig.~\ref{noi_vis_lin}, where the noise amplitudes are similar to those in the middle row of Fig.~\ref{noi_vis_std}. 
This behavior is reminiscent of the general fact that lensing signatures tend to appear at long baselines; for example, in the context of correlations in visibility space, \cite{wong24} found the lensing signatures to be manifested preferably at long baselines.

\begin{figure}[h!t]
\centering
\includegraphics[width=.75\textwidth]{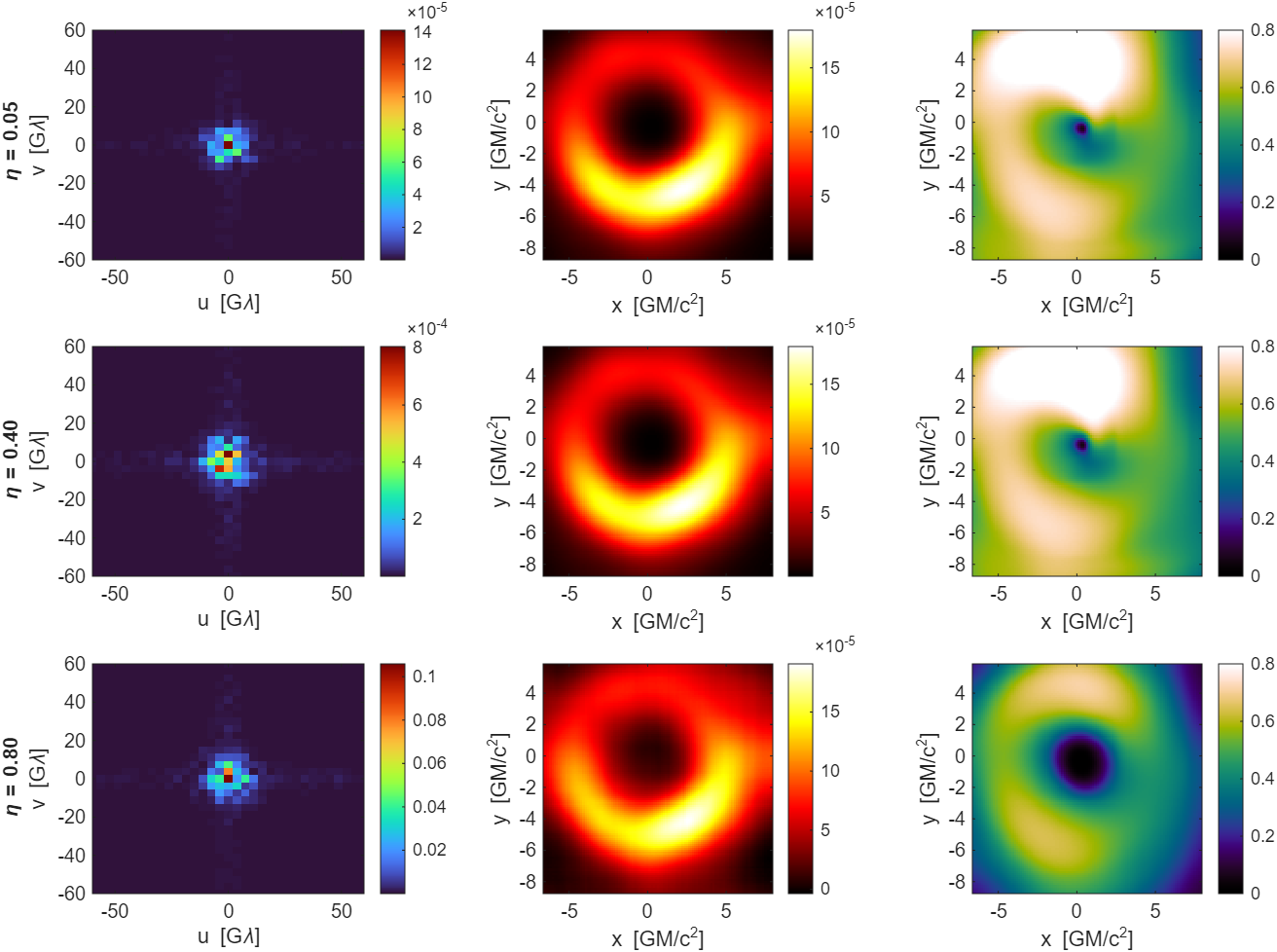}
\caption{Similar maps to those presented in Fig.~\ref{noi_vis_std}, but for the blurred movie with FWHM=10~$\mu$as.}\label{noi_vis_std_b10}
\end{figure}

\begin{figure}[h!t]
\centering
\includegraphics[width=.75\textwidth]{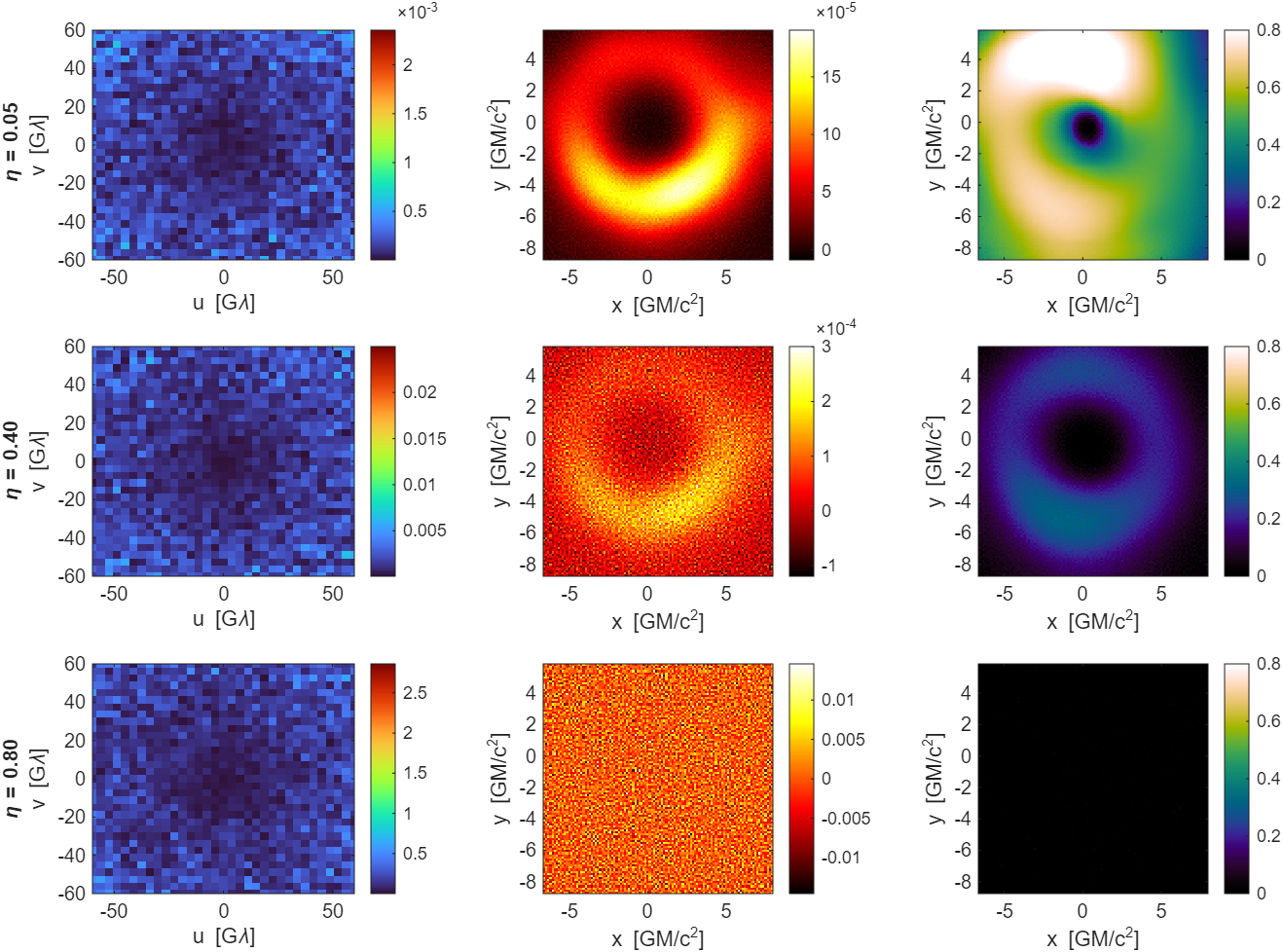}
\caption{Similar maps to those presented in Fig.~\ref{noi_vis_lin}, but for the blurred movie with FWHM=10~$\mu$as.}\label{noi_vis_lin_b10}
\end{figure}

The effect of the noise functions $\tilde{N}_1$, $\tilde{N}_2$ on the blurred movie is shown in Figs.~\ref{noi_vis_std_b10}, \ref{noi_vis_lin_b10}, respectively. The movie was blurred with a kernel of FWHM=10~$\mu$as. In general, the effect on correlation maps is consistent with the results from the unblurred movie, but the effects on individual snapshots are suppressed. The correlations are lower for noisier blurred movies, but since the correlations are overall higher in the blurred cases, the correlations with noise also remain higher compared to the unblurred movie. 
In the blurred movies, as may be expected due to the scaling with $\mathrm{std}\left[ |V(t,u,v)| \right]$, the noise function $\tilde{N}_1$ is even more localized around the origin in visibility space compared to the unblurred case, 
and hence the lensing correlation structure is even less influenced by such noise (Fig.~\ref{noi_vis_std_b10}). 

Injecting the noise $\tilde{N}_2$ to the blurred movie (Fig.~\ref{noi_vis_lin_b10}) introduces short-scale structure in the image plane, at scales that were suppressed before the noise was added.
This happens since the blurred movie is dominated by short baselines, while $\tilde{N}_2$ is dominated by long baselines.
This situation does not seem realistic for a VLBI measurement in which, by construction, both signal and noise are introduced at the same baselines. Nevertheless, we include this example for completeness;
in this case the lensing correlation structure seems rather sensitive to noise.

\section{Statistics of many fixed points}\label{sec_stat}
In the preceding sections, we have studied the two-point correlation structure and its sensitivity to various movie parameters for the case where one of the points is fixed. Here, we will initiate the discussion on how to exploit the full content of $\mathcal{C}$, or in other words, combine the information on different fixed points.
Developing a practical method to do so could provide a useful tool to eventually identify the photon ring and constrain the BH parameters 
(namely, spin and inclination) 
via spatiotemporal correlations in upcoming data. 

In this section we develop an approach that analyzes the statistics of the maxima associated with extreme lensing for a set of fixed points. 
For each fixed point, we characterize the lensing maximum by 
its time lag $T$ 
and deviation angle $\delta \varphi$. The latter indicates the angular deviation of the lensing peak from a shift of $\pi$ relative to the fixed point, expected for example in the zero inclination, zero spin case (see Sec.~\ref{sec_ident} for details).
After calculating the full 5D correlation function for all points on the observer screen, for all possible choices of fixed point, we computed the corresponding $T$ and $\delta \varphi$, as described in Sec.~\ref{sec_ident}. 
Having in mind a potential future observational application of this technique, we tried to make as few a priori assumptions as possible on the location of the maxima.

In order to avoid the false identification of correlation peaks that arise from fluctuations of the astrophysical correlation as lensing peaks,
in the following analysis we only included fixed points that have $\mathcal{C}\geq 0.5$ at their corresponding lensing peak. This threshold guarantees that only points with sufficient signal are considered, while keeping their number sufficiently high to perform reasonable statistics. Their exact number depends on the blurring, but typically several thousand points were included in each realization; at least 4,000~points in each case.

\begin{figure}[h!t]
\centering
\includegraphics[width=.85\textwidth]{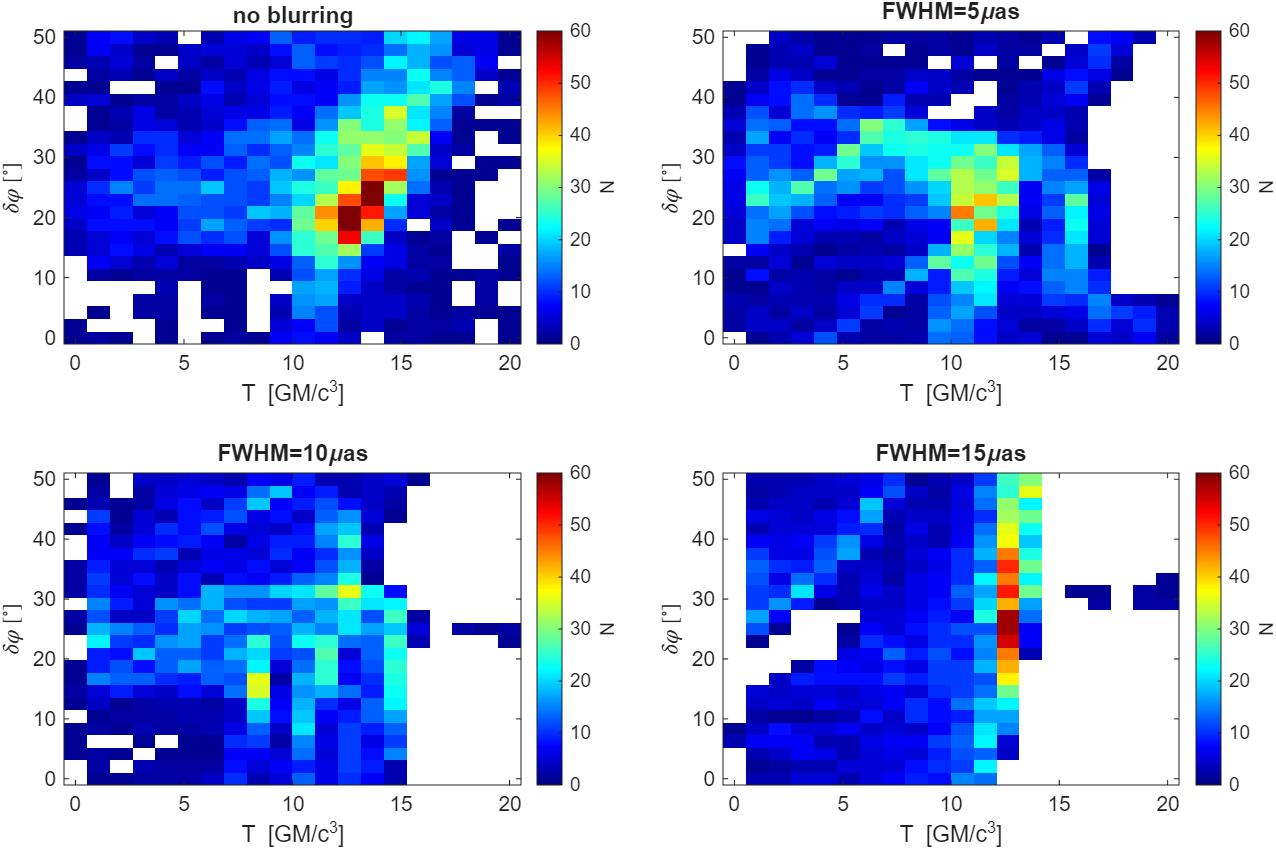}
\caption{2D histogram of the lensing maxima corresponding to different fixed points, 
derived from movies with different blurrings (individual panels), keeping only those that have $\mathcal{C}\geq 0.5$ at the peak. The maxima are binned by their values of $(T,\delta\varphi)$ (time lag and angular deviation from $\pi$, respectively).}\label{2d_hist}
\end{figure}

Fig.~\ref{2d_hist} shows a two-dimensional histogram displaying the number of above-threshold lensing maxima 
that fall in every (equally sized) bin in the $(T,\delta \varphi)$-plane, for different blurrings. 
The resulting distribution conveys valuable information on the source.
A robust feature, which seems to be quite stable across the range of different blurrings (different panels in Fig.~\ref{2d_hist}) is that the characteristic time lag of the distribution typically lies somewhere between 10 and 15~$GM/c^3$. As for deviation angles, we searched for maxima in the angular range $-\pi/3<\delta \varphi<\pi/3$, as discussed in Sec.~\ref{sec_ident}. We find that the characteristic angular deviation tends to be positive (we do not show here some rare negative values) and roughly in the range $15^\circ<\delta \varphi<35^\circ$.

\begin{figure}[h!t]
\centering
\includegraphics[width=.5\textwidth]{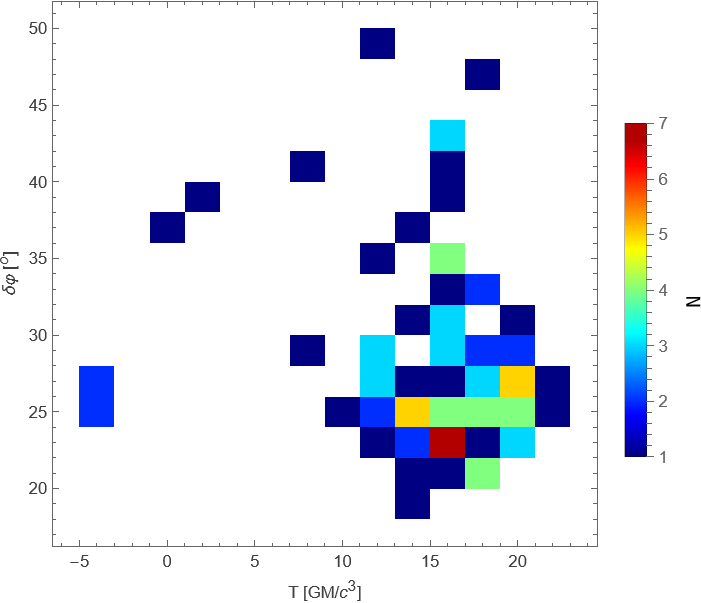}
\caption{2D histogram of pairs of $(T,\delta\varphi)$ which were theoretically computed for 160 arbitrarily selected equatorial point sources (the procedure is described in the body of the paper); cf. Fig.~\ref{2d_hist}.}\label{2d_hist_teor}
\end{figure}

Our findings can be compared with theoretical predictions which we 
developed using the methods introduced by \cite{zhou25}. We made a simplifying assumption that the source is equatorial, and for 160 points we chose on the observer screen we ray traced backwards towards the source, finding their intersection with the equatorial plane. Thinking of this intersection point as a point source, we \emph{forward ray trace} to find the $n=1$ ray reaching the observer, obtaining the first indirect image of that point source on the screen. The above procedure gives us pairs of $n=0$ (original point chosen on the screen) and $n=1$ images (inferred through the described procedure) of the same (equatorial) source points. 
Thus, we could generate a list of respective pairs $(T,\delta\varphi)$, presented in a histogram in Fig.~\ref{2d_hist_teor}. This provides a theoretical prediction for the distribution of lensing peaks if the sources of emission around a BH with $a=0.9375$, $\theta_0=163^\circ$, and $d=5.43\times10^{10}~MG/c^2$ (i.e., an M87*-like BH) are mostly equatorial. In particular, the characteristic $T$ should lie roughly between 10 and 20~$GM/c^3$ and the characteristic $\delta \varphi$ typically lies between $20^\circ$ and $35^\circ$.

It is important to stress that the overall characteristic lag and angular deviation convey only part of the information contained in the distribution. Its shape, whenever observationally accessible, can convey much more detailed information on the BH. In fact, even the shape of the distribution does not convey all lensing information, since for example it does not tell us which fixed point was associated with which point in the $(T,\delta\varphi)$-plane. Nevertheless, this method seems like a powerful tool to visualize and analyze correlation measurements.
It is also worth keeping in mind that the method described in this section is of course resolution-dependent. For lower resolution (here mimicked by larger blurring kernel FWHM) it might be challenging to obtain a sufficiently large number of independent data points.
Another noteworthy expectation is that at high inclination and significant spin, the distribution is expected to be much flatter, with no single characteristic lag or angular deviation, but a potentially richer shape in the $(T,\delta\varphi)$-plane to attempt to measure.
Moreover, there might be setbacks 
in some regions in the space of astrophysical parameters; namely, when the correlation lengths become large.
A detailed analysis of these questions would require generating more movies with different source parameters, and we defer it to future work.

Finally, even though in the BH movie analyzed here, our choice of the angular region to search for the maxima $-\pi/3<\delta \varphi<\pi/3$ has worked well, 
this, of course, does not have to be true for other values of BH and source parameters.
For different parameter choices, the possible values of $\delta \varphi$ may span different angular intervals, and this might create challenges in distinguishing the lensing and astrophysical peaks. However, the analysis of detailed limitations of our method along these lines is beyond the scope of this work. Within this study, we wanted to outline a new approach based on the statistics of many fixed points. 
This approach could initially serve to rule out some possible combinations of spin and inclination. Importantly, the characteristic $\delta\varphi$ and the distribution of angular deviations are unique observables which cannot be obtained from time-averaged images, even when the photon ring is spatially resolved. Correlations of a BH movie can hence provide independent pieces of information which, combined with other techniques, could contribute to the inference of observed BH properties, including its spin.

\section{Discussion and Conclusions}
In this study, we have expanded and elaborated our ongoing investigation of the following questions: is a spatiotemporal correlation analysis capable of uncovering signatures of general-relativistic extreme-lensing effects in BH movies to be captured by upcoming observations? What analysis methods can help us best extract, from the correlation structure, the physical information on the sources we observe? 
 
The movie---a time series of images---that we studied was produced by slow-light ray tracing of a GRMHD simulation, which was then corrupted in a variety of ways to mimic the expected limitations of upcoming realistic measurements. Following up on the analysis of \cite{bezdekova26}, our case study was a high-spin, low-inclination source (see Sec.~\ref{sec:simulated movie} for all source parameters) and we studied the two-point correlation function on a 3D subspace of the 5D configuration space, defined by fixing the earlier point on the observer's screen. This reduction allowed us to visualize the results and understand the correlation structure across the configuration space.

One of the key queries we aimed to address was under what conditions the correlation peak is identifiable, starting with the 3D reduced configuration subspace.  
The first realistic limitation we discuss is the movie's spatial resolution. Our analysis suggests that even a spatial resolution of 20~$\mu$as in itself does not preclude, at least in principle, an identification of the lensing correlation peak. 
Whether or not such an identification is possible in a particular setting depends also on other factors, including the choice of a fixed point, and possibly how limited the $(u,v)$-coverage used to reconstruct the movie is---see discussion below.

Regarding the minimal movie duration $L$ needed to reliably identify the lensing correlation peak, this value depends on the temporal correlation timescale of the source $\mathcal{K}$. The number of temporal statistical realizations in the movie may be roughly estimated as the ratio of these scales, and indeed we see that the correlation function approaches its long-duration limit as $\sim (L/\mathcal{K})^{-1/2}$, see Fig.~\ref{dur_point}.
For the movie we analyze here, that has total duration $L_\mathrm{tot}=4000 \, GM/c^3$, 
we show that a duration of $L\sim 2000~GM/c^3$ is enough for sufficient convergence of the correlation structure to its large-$L$ value. This estimate is encouraging when contemplating the specific target sources, M87* and Sgr A*, of future observation campaigns, where $2000~GM/c^3$ corresponds to $\sim$ 2 years and $\sim$ 11 h, respectively. However, whether or not such a duration will be enough depends on additional source and measurement parameters.
As long as the movie temporal resolution (cadence) is higher than a few snapshots per gravitational lensing timescale, $\sim15~GM/c^3$, it does not seem to be a limiting factor for correlation analysis. In fact, in Sec.~\ref{sec:cadence} we argue that a cadence of $\sim 1~\mathrm{min}$ should be sufficient for a correlation analysis of Sgr A*. For M87*, cadence is a non-issue because of the much longer timescale.

A significant issue that inevitably arises in real VLBI measurements is the temporal segmentation of the data, resulting from instrument calibration constraints due to Earth's rotation. This feature might cause false correlation peaks to occur and disrupt the identification of lensing signatures. However, as we have shown in Sec.~\ref{sec_seg}, this obstacle may be overcome if the pauses in data collection, between the segments, are short enough. In addition, it is preferable that the duration of segments is much longer than the characteristic lensing timescale $\sim15~GM/c^3$. If that is impossible, it seems preferable to make it much shorter than the lensing timescale so as not to inject artificial features at similar lags to the correlation function. This aspect is relevant for Sgr A* because (current) calibration timescales are comparable to its lensing timescale. It is also marginally relevant for M87*, where segmentation occurs not due to calibration issues, but directly due to Earth's rotation, as the stations can observe the target only for part of the day. However, since for M87* the lensing timescale is $15~GM/c^3\approx 5$ days, this segmentation is not expected to preclude a correlation measurement. 

Another inevitable aspect of real measurements is the presence of noise. In Sec.~\ref{sec_noise}, we studied several toy models of noise and their effect on correlation maps. We differentiated between noise functions that are naturally defined in the image/intensity domain with those that are natural from the point of view of the $(u,v)$-plane/visibility domain. 
Even though they are toy models, our noise functions give a general idea of the SNR levels that still allow an identification of the lensing signatures in correlation maps. Generally speaking, we can still see lensing signatures even with $\mathcal{O}(1)$ SNR, but this depends on additional aspects of the noise such as its spectral distribution.
We find that significant noise corruption at short baselines does not influence the correlations as much as comparable noise at long baselines, confirming a prior observation in the literature \citep{wong24}.
An important comment is in order: Adding noise directly to visibilities, when there is finite $(u,v)$-coverage, mimics the situation in real VLBI observations most closely. However, since we defer the treatment of finite $(u,v)$-coverage (combined with noise) to future work, the present noise analysis should be considered schematic.

At this stage, we would like to remark on visibility-domain correlation analyses. In principle, visibilities of particular baselines can be cross-correlated, in an analogous manner to cross-correlations of pixels in the intensity domain, as was done by 
\cite{wong24} who cross-correlated different baselines with the light curve (zero baseline). Correlating particular visibilities would definitely be the most straightforward procedure from the point of view of real VLBI measurements. 
Therefore, it would be desirable to find an optimal combination of visibilities that would display the extreme-lensing signatures in the clearest way. 
However, we have not been able to date to find a combination which displays the signatures more clearly than the 2D inverse Fourier transform itself: in other words, the image/intensity domain correlation function---as far as we are aware, it is the most effective combination, at least in the case where the source correlations are localized. 
This image-domain correlator may be computed with or without invoking advanced image processing techniques \citep[see,][]{EHT19IV,EHT22III} which are normally used by EHT for time-averaged images. Invoking these methods inevitably introduces biases into the analysis, while refraining from using them naturally yields ``dirty'' results. We hope to investigate these possibilities in the context of finite $(u,v)$-coverage effects in future work.

In Sec.~\ref{sec_stat}, we have briefly outlined a statistical approach to the analysis of the full 5D correlation function, which constitutes a different type of reduction compared to that performed in the previous sections. 
Even though $\mathcal{C}(T,x,y,x',y')$ contains an equal or larger amount of information on the source compared to any possible reduction of it (whether by projection, integration, or another method), extracting this information and deriving physical conclusions from it is challenging.
It is desirable to transcend the method of fixing an earlier point and analyzing the 3D projection of $\mathcal{C}$ manually in detail. We would like to exploit the fact that there are 
tens of thousands of points on the observer's screen.
In this approach, for each fixed point we automatically identify the time lag $T$ and angular deviation $\delta\varphi$, and we plot this distribution of these temporal-azimuthal `lags' as a 2D histogram, see Fig.~\ref{2d_hist}. We find that the shape of this distribution, including (but not limited to) the location of its peak, is a particularly useful visualization tool informing us about the lensing properties of the BH. This approach seems promising, but in order to assess its full utility a detailed study of more BH movies, with different spins and inclinations, is required.

It is worthwhile to briefly reiterate the main lessons and recommendations that emerge from our analysis of the present case study in the context of the two main BH-imaging targets. For M87* (the present movie parameters were chosen as best-guess parameters for this source), we find that cadence and data segmentation are non-issues, and the challenging aspect of observing its spatiotemporal correlations is the need for a continuous $\sim2$ year campaign. For Sgr A*, duration is a non-issue, and the challenge is to upgrade the cadence (including pauses for calibration) to a snapshot per $\sim 1$ minute or higher. For both sources, resolution and SNR---in themselves---do not seem to preclude an observation. Of course, it seems entirely possible that a combination of the above-mentioned realistic observational limitations would imply that longer duration/higher cadence, respectively, are required. 

In addition, as mentioned above, a substantial open question remains to be answered before a reliable simulated realistic test of a correlation measurement can be claimed. VLBI data only partially covers the $(u,v)$-plane, and usually (e.g., in EHT analysis) additional modeling techniques are used to complement the data. In the present work we did not touch this aspect at all. A crucial question which must be addressed in follow-up work is therefore how will the particular $(u,v)$-coverage of intended VLBI missions and (possibly) the respective modeling techniques used in image reconstruction affect the correlation functions? We leave this question for a separate study. 
Another crucial aspect is to generalize the present case study by analyzing more movies: both for different choices of BH parameters---spin and inclination---and for different choices of astrophysical parameters characterizing the emitting plasma.

We hope that our work provides a better understanding of how correlations of image fluctuations are affected by realistic limitations, and supports the hypothesis that they can serve as a complementary tool for probing BHs.

\begin{acknowledgments}
We are grateful to G.N. Wong for providing us with the black-hole movie analyzed in this study and to A. C{\'a}rdenas-Avenda{\~n}o for comments on the manuscript. This work was supported in part by the Israel Science Foundation (grant no. 2047/23) and the Binational Science Foundation (grant no. 024239). B.B. acknowledges support from the Simons Foundation
(MP-SCMPS-00001470). The computations that led to the results
presented in this work were partially performed on the Hive computer
cluster at the University of Haifa, which is partly funded by ISF
grant 2155/15.
\end{acknowledgments}

\begin{contribution}
B.B. and S.H. formulated the structure of the paper. B.B. performed the analysis and elaborated on the corresponding techniques. B.B. and S.H. wrote the paper.
\end{contribution}

\bibliography{sample701}{}

@ARTICLE{hadar21,
       author = {{Hadar}, Shahar and {Johnson}, Michael D. and {Lupsasca}, Alexandru and {Wong}, George N.},
        title = "{Photon ring autocorrelations}",
      journal = "Phys. Rev. D",
         year = "2021",
        month = "May",
       volume = {103},
       number = {10},
          eid = {104038},
        pages = {104038},
          doi = {10.1103/PhysRevD.103.104038},
       adsurl = {https://ui.adsabs.harvard.edu/abs/2021PhRvD.103j4038H},
}

@ARTICLE{Chael2021,
       author = {{Chael}, Andrew and {Johnson}, Michael D. and {Lupsasca}, Alexandru},
        title = "{Observing the Inner Shadow of a Black Hole: A Direct View of the Event Horizon}",
      journal = {\apj},
         year = 2021,
        month = sep,
       volume = {918},
       number = {1},
          eid = {6},
        pages = {6},
          doi = {10.3847/1538-4357/ac09ee},
archivePrefix = {arXiv},
       eprint = {2106.00683},
 primaryClass = {astro-ph.HE},
       adsurl = {https://ui.adsabs.harvard.edu/abs/2021ApJ...918....6C}
}

@article{andrianov22,
       author = {Andrianov, A and others},
        title = {Flares and their echoes can help distinguish photon rings from black holes with space-{E}arth very long baseline interferometry},
      journal = {Phys. Rev. D},
       volume = {105},
       number = {6},
        pages = {063015},
         year = {2022},
    publisher = {APS},
         doi = {10.1103/PhysRevD.105.063015}
}

@article{aratore24,
       author = {Aratore, Fabio and Tsupko, Oleg Yu and Perlick, Volker},
        title = {Constraining spherically symmetric metrics by the gap between photon rings},
      journal = {Physical Review D},
       volume = {109},
       number = {12},
        pages = {124057},
         year = {2024},
    publisher = {APS},
          doi = {10.1103/PhysRevD.109.124057}
}

@INPROCEEDINGS{bardeen73,
       author = {{Bardeen}, J.~M.},
        title = "{Timelike and null geodesics in the Kerr metric.}",
    booktitle = {Black Holes (Les Astres Occlus)},
         year = 1973,
       editor = {{Dewitt}, C. and {Dewitt}, B.~S.},
        month = jan,
        pages = {215-239},
       adsurl = {https://ui.adsabs.harvard.edu/abs/1973blho.conf..215B}
}

@article{beckwith05,
       author = {Beckwith, Kris and Done, Chris},
        title = {Extreme gravitational lensing near rotating black holes},
      journal = {Monthly Notices of the Royal Astronomical Society},
       volume = {359},
       number = {4},
        pages = {1217--1228},
         year = {2005},
          doi = {10.1111/j.1365-2966.2005.08980.x},
    publisher = {Blackwell Science Ltd Oxford, UK}
}

@article{bezdekova26,
       author = {Bezd{\v{e}}kov{\'a}, Barbora and Hadar, Shahar and Wong, George N and Wielgus, Maciek},
        title = {Extreme lensing signatures revealed by correlations of simulated black hole movies},
      journal = {Nature Astronomy},
        pages = {1199--1207},
       volume = {10},
       number = {8},
         year = {2026},
          doi = {10.1038/s41550-026-02874-x},
    publisher = {Nature Publishing Group UK London}
}

@INPROCEEDINGS{BHEX2024,
       author = {{Johnson}, Michael D. and others},
        title = "{The Black Hole Explorer: motivation and vision}",
    booktitle = {Space Telescopes and Instrumentation 2024: Optical, Infrared, and Millimeter Wave},
         year = 2024,
       editor = {{Coyle}, Laura E. and {Matsuura}, Shuji and {Perrin}, Marshall D.},
       series = {Society of Photo-Optical Instrumentation Engineers (SPIE) Conference Series},
       volume = {13092},
        month = aug,
          eid = {130922D},
        pages = {130922D},
          doi = {10.1117/12.3019835},
       adsurl = {https://ui.adsabs.harvard.edu/abs/2024SPIE13092E..2DJ}
}

@article{blackburn20,
       author = {Blackburn, Lindy and Pesce, Dominic W and Johnson, Michael D and Wielgus, Maciek and Chael, Andrew A and Christian, Pierre and Doeleman, Sheperd S},
        title = {Closure statistics in interferometric data},
      journal = {The Astrophysical Journal},
       volume = {894},
       number = {1},
        pages = {31},
         year = {2020},
          doi = {10.3847/1538-4357/ab8469},
    publisher = {The American Astronomical Society}
}

@article{cardenas24,
       author = {C{\'a}rdenas-Avenda{\~n}o, Alejandro and Gammie, Charles and Lupsasca, Alexandru},
        title = {Explanation for the absence of secondary peaks in black hole light curve autocorrelations},
      journal = "Phys. Rev. Lett.",
       volume = {133},
       number = {13},
        pages = {131402},
         year = {2024},
          doi = {10.1103/PhysRevLett.133.131402},
    publisher = {APS}
}

@ARTICLE{Carter1968,
       author = {{Carter}, Brandon},
        title = "{Global Structure of the Kerr Family of Gravitational Fields}",
      journal = {Phys. Rev.},
         year = 1968,
        month = oct,
       volume = {174},
       number = {5},
        pages = {1559-1571},
          doi = {10.1103/PhysRev.174.1559},
       adsurl = {https://ui.adsabs.harvard.edu/abs/1968PhRv..174.1559C}
}

@ARTICLE{cunningham73,
       author = {{Cunningham}, C.~T. and {Bardeen}, James M.},
        title = "{The Optical Appearance of a Star Orbiting an Extreme Kerr Black Hole}",
      journal = {The Astrophysical Journal},
         year = 1973,
        month = jul,
       volume = {183},
        pages = {237-264},
          doi = {10.1086/152223},
       adsurl = {https://ui.adsabs.harvard.edu/abs/1973ApJ...183..237C},
}

@ARTICLE{Doeleman2023,
       author = {{Doeleman}, Sheperd S. and others},
        title = "{Reference Array and Design Consideration for the Next-Generation Event Horizon Telescope}",
      journal = {Galaxies},
         year = 2023,
        month = oct,
       volume = {11},
       number = {5},
          eid = {107},
        pages = {107},
          doi = {10.3390/galaxies11050107},
       adsurl = {https://ui.adsabs.harvard.edu/abs/2023Galax..11..107D}
}

@article{edelson88,
       author = {Edelson, Rick A and Krolik, Julian H},
        title = {The discrete correlation function-A new method for analyzing unevenly sampled variability data},
      journal = {Astrophys. J.},
       volume = {333},
        pages = {646--659},
         year = {1988},
         doi = {10.1086/166773}
}

@article{EHT19I,
       author = {{The Event Horizon Telescope Collaboration} and others},
        title = {First {M}87 {E}vent {H}orizon {T}elescope {R}esults. {I}. {T}he {S}hadow of the {S}upermassive {B}lack {H}ole},
      journal = {Astrophys. J. Lett.},
       volume = {875},
        pages = {L1},
         year = {2019},
          doi = {10.3847/2041-8213/ab0ec7},
    publisher = {IOP Publishing}
}

@article{eht19ii,
       author = {{The Event Horizon Telescope Collaboration} and others},
        title = {First {M87} {E}vent {H}orizon {T}elescope results. {II}. {A}rray and instrumentation},
      journal = {The Astrophysical Journal Letters},
       volume = {875},
       number = {1},
        pages = {L2},
         year = {2019},
          doi = {10.3847/2041-8213/ab0c96},
    publisher = {The American Astronomical Society}
}

@article{eht19iii,
       author = {{The Event Horizon Telescope Collaboration} and others},
        title = {First {M87} {E}vent {H}orizon {T}elescope results. {III}. {D}ata processing and calibration},
      journal = {The Astrophysical Journal Letters},
       volume = {875},
       number = {1},
        pages = {L3},
         year = {2019},
          doi = {10.3847/2041-8213/ab0c57},
    publisher = {IoP Publishing}
}

@ARTICLE{EHT19IV,
       author = {{The Event Horizon Telescope Collaboration} and others},
        title = "{First {M87} {E}vent {H}orizon {T}elescope Results. {IV}. {I}maging the Central Supermassive Black Hole}",
      journal = {Astrophys. J. Lett.},
         year = 2019,
        month = apr,
       volume = {875},
       number = {1},
          eid = {L4},
        pages = {L4},
          doi = {10.3847/2041-8213/ab0e85},
       adsurl = {https://ui.adsabs.harvard.edu/abs/2019ApJ...875L...4E}
}

@ARTICLE{EHT19V,
       author = {{The Event Horizon Telescope Collaboration} and others},
        title = "{First {M87} {E}vent {H}orizon {T}elescope Results. {V}. {P}hysical Origin of the Asymmetric Ring}",
      journal = {Astrophys. J. Lett.},
         year = 2019,
        month = apr,
       volume = {875},
       number = {1},
          eid = {L5},
        pages = {L5},
          doi = {10.3847/2041-8213/ab0f43},
       adsurl = {https://ui.adsabs.harvard.edu/abs/2019ApJ...875L...5E},
}

@ARTICLE{EHT19VI,
       author = {{The Event Horizon Telescope Collaboration} and others},
        title = "{First {M87} {E}vent {H}orizon {T}elescope Results. {VI}. {T}he Shadow and Mass of the Central Black Hole}",
      journal = {Astrophys. J.},
         year = 2019,
        month = apr,
       volume = {875},
       number = {1},
          eid = {L6},
        pages = {L6},
          doi = {10.3847/2041-8213/ab1141},
       adsurl = {https://ui.adsabs.harvard.edu/abs/2019ApJ...875L...5E},
}

@ARTICLE{EHT22I,
       author = {{The Event Horizon Telescope Collaboration} and others},
        title = "{First {S}agittarius {A}* {E}vent {H}orizon {T}elescope {R}esults. {I}. {T}he Shadow of the Supermassive Black Hole in the Center of the Milky Way}",
      journal = {Astrophys. J. Lett.},
         year = 2022,
        month = may,
       volume = {930},
       number = {2},
          eid = {L12},
        pages = {L12},
          doi = {10.3847/2041-8213/ac6674},
       adsurl = {https://ui.adsabs.harvard.edu/abs/2022ApJ...930L..12E},
}

@ARTICLE{EHT22III,
       author = {{The Event Horizon Telescope Collaboration} and others},
        title = "{First Sagittarius A* Event Horizon Telescope Results. III. Imaging of the Galactic Center Supermassive Black Hole}",
      journal = {Astrophys. J. Lett.},
         year = 2022,
        month = may,
       volume = {930},
       number = {2},
          eid = {L14},
        pages = {L14},
          doi = {10.3847/2041-8213/ac6429},
       adsurl = {https://ui.adsabs.harvard.edu/abs/2022ApJ...930L..14E},
}

@INPROCEEDINGS{falcke17,
       author = {{Falcke}, Heino},
        title = "{Imaging black holes: past, present and future}",
    booktitle = {Journal of Physics Conference Series},
         year = 2017,
       series = {Journal of Physics Conference Series},
       volume = {942},
        month = dec,
    publisher = {IOP},
          eid = {012001},
        pages = {012001},
          doi = {10.1088/1742-6596/942/1/012001},
}

@article{fukumura08,
       author = {Fukumura, Keigo and Kazanas, Demosthenes},
        title = {Light echoes in {K}err geometry: A source of high-frequency {QPO}s from random {X}-ray bursts},
      journal = {Astrophys. J.},
       volume = {679},
       number = {2},
        pages = {1413},
         year = {2008},
          doi = {10.1086/587159},
    publisher = {IOP Publishing}
}

@ARTICLE{gralla20lensing,
       author = {{Gralla}, Samuel E. and {Lupsasca}, Alexandru},
        title = "{Lensing by Kerr black holes}",
      journal = {Phys. Rev. D},
         year = 2020,
        month = feb,
       volume = {101},
       number = {4},
          eid = {044031},
        pages = {044031},
          doi = {10.1103/PhysRevD.101.044031},
       adsurl = {https://ui.adsabs.harvard.edu/abs/2020PhRvD.101d4031G}
}

@ARTICLE{Gralla2019,
       author = {{Gralla}, Samuel E. and {Holz}, Daniel E. and {Wald}, Robert M.},
        title = "{Black hole shadows, photon rings, and lensing rings}",
      journal = {Phys. Rev. D},
         year = 2019,
        month = jul,
       volume = {100},
       number = {2},
          eid = {024018},
        pages = {024018},
          doi = {10.1103/PhysRevD.100.024018},
       adsurl = {https://ui.adsabs.harvard.edu/abs/2019PhRvD.100b4018G}
}

@article{hadar23,
       author = {Hadar, Shahar and Harikesh, Sreehari and Chelouche, Doron},
        title = {Extreme lensing induces spectrotemporal correlations in black-hole signals},
      journal = {Phys. Rev. D},
       volume = {107},
       number = {12},
        pages = {124057},
         year = {2023},
          doi = {10.1103/PhysRevD.107.124057},
    publisher = {APS}
}

@article{harikesh25,
       author = {Harikesh, Sreehari and Hadar, Shahar and Chelouche, Doron},
        title = {Exploring lensing signatures through spectrotemporal correlations: Implications for black hole parameter estimation},
      journal = {Phys. Rev. D},
       volume = {112},
       number = {4},
        pages = {043020},
         year = {2025},
          doi = {10.1103/wjpm-9byt},
    publisher = {APS}
}

@article{jia24,
       author = {Jia, He and Quataert, Eliot and Lupsasca, Alexandru and Wong, George N},
        title = {Photon ring interferometric signatures beyond the universal regime},
      journal = {Phys. Rev. D},
       volume = {110},
       number = {8},
        pages = {083044},
         year = {2024},
          doi = {10.1103/PhysRevD.110.083044},
    publisher = {APS}
}

@ARTICLE{Johnson2020,
       author = {{Johnson}, Michael D. and others},
        title = "{Universal interferometric signatures of a black hole's photon ring}",
      journal = {Sci. Adv.},
         year = 2020,
        month = mar,
       volume = {6},
       number = {12},
        pages = {eaaz1310},
          doi = {10.1126/sciadv.aaz1310},
       adsurl = {https://ui.adsabs.harvard.edu/abs/2020SciA....6.1310J}
}

@article{johnson23,
       author = {Johnson, Michael D and others},
        title = {Key science goals for the next-generation {E}vent {H}orizon {T}elescope},
      journal = {Galaxies},
       volume = {11},
       number = {3},
        pages = {61},
         year = {2023},
          doi = {10.3390/galaxies11030061},
    publisher = {MDPI}
}

@ARTICLE{luminet79,
       author = {{Luminet}, J. -P.},
        title = "{Image of a spherical black hole with thin accretion disk.}",
      journal = {Astronomy and Astrophysics},
         year = 1979,
        month = may,
       volume = {75},
        pages = {228-235},
       adsurl = {https://ui.adsabs.harvard.edu/abs/1979A&A....75..228L},
}

@incollection{lupsasca24,
       author = {Lupsasca, Alexandru and Mayerson, Daniel R and Ripperda, Bart and Staelens, Seppe},
        title = {A beginner’s guide to black hole imaging and associated tests of general relativity},
    booktitle = {Recent Progress on Gravity Tests: Challenges and Future Perspectives},
        pages = {183--237},
         year = {2024},
          doi = {10.1007/978-981-97-2871-8_6},
    publisher = {Springer} 
}

@article{medeiros23,
       author = {Medeiros, Lia and Psaltis, Dimitrios and Lauer, Tod R and {\"O}zel, Feryal},
        title = {Principal-component interferometric modeling ({PRIMO}), an algorithm for {EHT} data. {I}. {R}econstructing images from simulated {EHT} observations},
      journal = {The Astrophysical Journal},
       volume = {943},
       number = {2},
        pages = {144},
         year = {2023},
          doi = {10.3847/1538-4357/acaa9a},
    publisher = {The American Astronomical Society} 
}

@article{moriyama19,
       author = {Moriyama, Kotaro and Mineshige, Shin and Honma, Mareki and Akiyama, Kazunori},
        title = {Black hole spin measurement based on time-domain {VLBI} observations of infalling gas clouds},
      journal = {Astrophys. J.},
       volume = {887},
       number = {2},
        pages = {227},
         year = {2019},
          doi = {https://doi.org/10.3847/1538-4357/ab505b},
    publisher = {IOP Publishing}
}

@article{moscibrodzka18,
       author = {{Mo{\'s}cibrodzka}, M and {Gammie}, Charles F},
        title = {ipole--semi-analytic scheme for relativistic polarized radiative transport},
      journal = {Mon. Not. R. Astron. Soc.},
       volume = {475},
       number = {1},
        pages = {43--54},
         year = {2018},
          doi = {10.1093/mnras/stx3162},
    publisher = {Oxford University Press}
}

@ARTICLE{Moscibrodzka2016,
       author = {{Mo{\'s}cibrodzka}, Monika and {Falcke}, Heino and {Shiokawa}, Hotaka},
        title = "{General relativistic magnetohydrodynamical simulations of the jet in M 87}",
      journal = {Astron. Astrophys.},
         year = 2016,
        month = feb,
       volume = {586},
          eid = {A38},
        pages = {A38},
          doi = {10.1051/0004-6361/201526630},
       adsurl = {https://ui.adsabs.harvard.edu/abs/2016A&A...586A..38M}
}

@article{ozel22,
       author = {{\"O}zel, Feryal and Psaltis, Dimitrios and Younsi, Ziri},
        title = {Black hole images as tests of general relativity: effects of plasma physics},
      journal = {The Astrophysical Journal},
       volume = {941},
       number = {1},
        pages = {88},
         year = {2022},
          doi = {10.3847/1538-4357/ac9fcb},
    publisher = {The American Astronomical Society}
}

@article{perlick22,
       author = {Perlick, Volker and Tsupko, Oleg Yu},
        title = {Calculating black hole shadows: {R}eview of analytical studies},
      journal = {Physics Reports},
       volume = {947},
        pages = {1--39},
         year = {2022},
          doi = {10.1016/j.physrep.2021.10.004},
    publisher = {Elsevier}
}

@article{pesce24,
       author = {Pesce, Dominic W and Blackburn, Lindy and Chaves, Ryan and Doeleman, Sheperd S and Freeman, Mark and Issaoun, Sara and Johnson, Michael D and Lindahl, Greg and Natarajan, Iniyan and Paine, Scott N and Palumbo, Daniel C M and Roelofs, Freek and Tiede, Paul},
        title = {Atmospheric limitations for high-frequency ground-based very long baseline interferometry},
      journal = {The Astrophysical Journal},
       volume = {968},
       number = {2},
        pages = {69},
         year = {2024},
          doi = {10.3847/1538-4357/ad3961},         
    publisher = {The American Astronomical Society}
}

@ARTICLE{prather21,
       author = {{Prather}, Ben and {Wong}, George and {Dhruv}, Vedant and {Ryan}, Benjamin and {Dolence}, Joshua and {Ressler}, Sean and {Gammie}, Charles},
        title = "{iharm3D: Vectorized General Relativistic Magnetohydrodynamics}",
      journal = {The Journal of Open Source Software},
         year = 2021,
        month = oct,
       volume = {6},
       number = {66},
          eid = {3336},
        pages = {3336},
          doi = {10.21105/joss.03336},
archivePrefix = {arXiv},
       eprint = {2110.10191},
 primaryClass = {astro-ph.HE},
       adsurl = {https://ui.adsabs.harvard.edu/abs/2021JOSS....6.3336P}
}

@article{rees82,
       author = {Rees, M. J. and Begelman, M. C. and Blandford, R. D. and Phinney, E. S.},
        title = {Ion-supported tori and the origin of radio jets},
      journal = {Nature},
       volume = {295},
       number = {5844},
        pages = {17--21},
         year = {1982},
          doi = {10.1038/295017a0},         
    publisher = {Nature Publishing Group UK London}
}

@article{reynolds96,
       author = {Reynolds, C. S. and Di Matteo, T. and Fabian, A. C. and Hwang, U. and Canizares, C. R.},
        title = {The ‘quiescent’ black hole in {M87}},
      journal = {Monthly Notices of the Royal Astronomical Society},
       volume = {283},
       number = {4},
        pages = {L111--L116},
         year = {1996},
          doi = {10.1093/mnras/283.4.L111},         
    publisher = {Blackwell Science Ltd Oxford, UK}
}

@inproceedings{sridharan24,
       author = {Sridharan, T. K. and others},
        title = {The black hole explorer ({BHEX}): Preliminary antenna design},
    booktitle = {Space Telescopes and Instrumentation 2024: Optical, Infrared, and Millimeter Wave},
       volume = {13092},
        pages = {2205--2211},        
         year = {2024},
          doi = {10.1117/12.3020504}
}

@article{tiede20,
       author = {Tiede, Paul and Pu, Hung-Yi and Broderick, Avery E and Gold, Roman and Karami, Mansour and Preciado-L{\'o}pez, Jorge A},
        title = {Spacetime tomography using the {E}vent {H}orizon {T}elescope},
      journal = {The Astrophysical Journal},
       volume = {892},
       number = {2},
        pages = {132},
         year = {2020},
          doi = {10.3847/1538-4357/ab744c},
    publisher = {The American Astronomical Society} 
}

@article{viergutz93,
       author = {Viergutz, S. U.},
        title = {Image generation in {K}err geometry. {I}. {A}nalytical investigations on the stationary emitter-observer problem},
      journal = {Astronomy and Astrophysics},
       volume = {272},
        pages = {355-377},
         year = {1993} 
}

@ARTICLE{Vincent2022,
       author = {{Vincent}, F.~H. and {Gralla}, S.~E. and {Lupsasca}, A. and {Wielgus}, M.},
        title = "{Images and photon ring signatures of thick disks around black holes}",
      journal = {Astron. Astrophys.},
         year = 2022,
        month = nov,
       volume = {667},
          eid = {A170},
        pages = {A170},
          doi = {10.1051/0004-6361/202244339},
       adsurl = {https://ui.adsabs.harvard.edu/abs/2022A&A...667A.170V}
}

@article{wielgus22,
       author = {Wielgus, Maciek and others},
        title = {Millimeter light curves of {S}agittarius {A}* observed during the 2017 {E}vent {H}orizon {T}elescope campaign},
      journal = {Astrophys. J. Lett.},
       volume = {930},
       number = {2},
        pages = {L19},
         year = {2022},
          doi = {10.3847/2041-8213/ac6428},
    publisher = {IOP Publishing}
}

@article{wong24,
       author = {Wong, George N and Medeiros, Lia and C{\'a}rdenas-Avenda{\~n}o, Alejandro and Stone, James M},
        title = {Measuring Black Hole Light Echoes with Very Long Baseline Interferometry},
      journal = {Astrophys. J. Lett.},
       volume = {975},
       number = {2},
        pages = {L40},
         year = {2024},
          doi = {10.3847/2041-8213/ad8650},
    publisher = {IOP Publishing}
}

@article{zhang25,
       author = {Zhang, Zhenyu and Hou, Yehui and Guo, Minyong and Mizuno, Yosuke and Chen, Bin},
        title = {Autocorrelation signatures in time-resolved black hole flare images: {S}econdary peaks and convergence structure},
      journal = {Physical Review D},
       volume = {112},
       number = {8},
        pages = {083024},
         year = {2025},
          doi = {10.1103/zmnz-p2rs},
    publisher = {APS}
}

@article{zhou25,
       author = {Zhou, Lihang and Zhong, Zhen and Chen, Yifan and Cardoso, Vitor},
        title = {Forward ray tracing and hot spots in {K}err spacetime},
      journal = {Phys. Rev. D},
       volume = {111},
       number = {6},
        pages = {064075},
         year = {2025},
          doi = {10.1103/PhysRevD.111.064075},
    publisher = {APS}
}

@ARTICLE{Gurvits2022,
       author = {{Gurvits}, Leonid I. and others},
        title = "{The science case and challenges of space-borne sub-millimeter interferometry}",
      journal = {Acta Astronaut.},
         year = 2022,
        month = jul,
       volume = {196},
        pages = {314-333},
          doi = {10.1016/j.actaastro.2022.04.020},
       adsurl = {https://ui.adsabs.harvard.edu/abs/2022AcAau.196..314G}
}

@ARTICLE{Hudson2023,
       author = {{Hudson}, Ben and others},
        title = "{Orbital configurations of spaceborne interferometers for studying photon rings of supermassive black holes}",
      journal = {Acta Astronaut.},
         year = 2023,
        month = dec,
       volume = {213},
        pages = {681-693},
          doi = {10.1016/j.actaastro.2023.09.035},
       adsurl = {https://ui.adsabs.harvard.edu/abs/2023AcAau.213..681H}
}

@ARTICLE{Urso2025,
       author = {{Urso}, I. and {Vincent}, F.~H. and {Wielgus}, M. and {Paumard}, T. and {Perrin}, G.},
        title = "{Gravity versus astrophysics in black hole images and photon rings: Equatorial emissions and spherically symmetric space-times}",
      journal = {Astron. Astrophys.},
         year = 2025,
        month = aug,
       volume = {700},
          eid = {A193},
        pages = {A193},
          doi = {10.1051/0004-6361/202554919},
       adsurl = {https://ui.adsabs.harvard.edu/abs/2025A&A...700A.193U}
}
\bibliographystyle{aasjournalv7}


\end{document}